\documentclass[10pt,twoside]{article}
\usepackage{graphicx}
\usepackage{amsmath}
\usepackage{Latex-document}

\def\volumeno{I}

\title{\bf Mathematical Foundations of \vskip -2mm Modern Cryptography: \vskip -2mm
Computational Complexity Perspective\vskip 6mm}

\author{Shafi Goldwasser\thanks{Department of Computer Science and
Applied Mathematics, Weizmann Institute, Israel and Department of
Electrical Engineering and Computer Science, Massachusetts
Institute of Technology, USA. E-mail: shafi@theory.lcs.mit.edu}
\vspace*{-0.5cm}}

\date{\vspace{-8mm}}

\newlength{\saveparindent}
\newlength{\saveparskip}

\newif\ifshortconferences
\shortconferencesfalse

\def\ending#1{{\count1=#1\relax
\count2=\count1 \divide\count2 by 100 \multiply\count2 by 100
\advance\count1 by -\count2
\ifnum\count1=11
th%
\else \ifnum\count1=12
th%
\else \ifnum\count1=13
th%
\else \count2=\count1 \divide\count1 by 10 \multiply\count1 by 10
\advance\count2 by -\count1 \ifnum\count2=1
st%
\else \ifnum\count2=2
nd%
\else \ifnum\count2=3
rd%
\else th%
\fi\fi\fi\fi\fi\fi }}

\def\Proceedings{\ifshortconferences {\sl Proc.}\else {\sl Proceedings}\fi}

\newcounter{confnum}

\def\conf#1#2{%
\setcounter{confnum}{#2}%
\addtocounter{confnum}{-\csname #1zero\endcsname}%
\ifnum\value{confnum}=1%
\expandafter\ifx\csname #1One\endcsname\relax%
\Proceedings\ {\sl of the}
\arabic{confnum}\ending{\value{confnum}}\ \csname #1name
\endcsname,\ \csname #1pub\endcsname,\ 19#2%
\else \csname #1One\endcsname\fi%
\else\ \Proceedings\ {\sl of the}
\arabic{confnum}\ending{\value{confnum}}\ \csname #1name\endcsname,\ %
\csname #1pub\endcsname,\ 19#2\fi}

\newcommand{\istcs}[1]{\ifnum#1=
93{{\sl Proceedings of the Second Israel Symposium on Theory and
Computing Systems\/}, IEEE, 1993}\else{\ifnum#1= 95{{\sl
Proceedings of the Third Israel Symposium on Theory and Computing
Systems\/}, IEEE, 1995}\else This ISTCS not yet defined! \fi}\fi}

\def\svconf#1#2{%
\csname #1name\endcsname~#2 {\sl Proceedings}, Lecture Notes in
Computer Science Vol.~\csname #1vol\endcsname{#2}, \csname
#1ed\endcsname{#2} ed., Springer-Verlag, 19#2}

\def\CRYPTOvol#1{\ifnum#1=
84{196}\else{\ifnum#1= 85{218}\else{\ifnum#1=
86{263}\else{\ifnum#1= 87{293}\else{\ifnum#1=
88{403}\else{\ifnum#1= 89{435}\else{\ifnum#1=
90{537}\else{\ifnum#1= 91{576}\else{\ifnum#1=
92{740}\else{\ifnum#1= 93{773}\else{\ifnum#1=
94{839}\else{\ifnum#1= 95{963}\else{\ifnum#1=
96{1109}\else{\ifnum#1= 97{1294}\else{\ifnum#1=
98{1462}\else{\ifnum#1=
99{??}\fi}\fi}\fi}\fi}\fi}\fi}\fi}\fi}\fi}\fi}\fi}\fi}\fi}\fi}\fi}\fi}

\def\CRYPTOed#1{\ifnum#1=
84{R.~Blakely}\else{\ifnum#1= 85{H.~Williams}\else{\ifnum#1=
86{A.~Odlyzko}\else{\ifnum#1= 87{C.~Pomerance}\else{\ifnum#1=
88{S.~Goldwasser}\else{\ifnum#1= 89{G.~Brassard}\else{\ifnum#1=
90{A.~J.~Menezes and S.~Vanstone}\else{\ifnum#1=
91{J.~Feigenbaum}\else{\ifnum#1= 92{E.~Brickell}\else{\ifnum#1=
93{D.~Stinson}\else{\ifnum#1= 94{Y.~Desmedt}\else{\ifnum#1=
95{D.~Coppersmith}\else{\ifnum#1= 96{N.~Koblitz}\else{\ifnum#1=
97{B.~Kaliski}\else{\ifnum#1= 98{H.~Krawczyk}\else{\ifnum#1=
99{M.~Wiener}\fi}\fi}\fi}\fi}\fi}\fi}\fi}\fi}\fi}\fi}\fi}\fi}\fi}\fi}\fi}\fi}

\def\EUROCRYPTvol#1{\ifnum#1=
84{209}\else{\ifnum#1= 85{219}\else{\ifnum#1=
86{??}\else{\ifnum#1=   
87{304}\else{\ifnum#1= 88{330}\else{\ifnum#1=
89{434}\else{\ifnum#1= 90{473}\else{\ifnum#1=
91{547}\else{\ifnum#1= 92{658}\else{\ifnum#1=
93{765}\else{\ifnum#1= 94{950}\else{\ifnum#1=
95{921}\else{\ifnum#1= 96{1070}\else{\ifnum#1=
97{1233}\else{\ifnum#1= 98{1403}\else{\ifnum#1=
99{??}\fi}\fi}\fi}\fi}\fi}\fi}\fi}\fi}\fi}\fi}\fi}\fi}\fi}\fi}\fi}\fi}

\def\EUROCRYPTed#1{\ifnum#1=
84{T.~Beth}\else{\ifnum#1= 85{F.~Pichler}\else{\ifnum#1=
86{??}\else{\ifnum#1=    
87{D.~Chaum}\else{\ifnum#1= 88{C.~Gunther}\else{\ifnum#1=
89{J-J.~Quisquater, J.~Vandewille}\else{\ifnum#1=
90{I.~Damg{\aa}rd}\else{\ifnum#1= 91{D.~Davies}\else{\ifnum#1=
92{R.~Rueppel}\else{\ifnum#1= 93{T.~Helleseth}\else{\ifnum#1=
94{A.~De~Santis}\else{\ifnum#1= 95{L.~Guillou and
J.~Quisquater}\else{\ifnum#1= 96{U.~Maurer}\else{\ifnum#1=
97{W.~Fumy}\else{\ifnum#1= 98{K.~Nyberg}\else{\ifnum#1=
99{J.~Stern}\fi}\fi}\fi}\fi}\fi}\fi}\fi}\fi}\fi}\fi}\fi}\fi}\fi}\fi}\fi}\fi}

\def\AUSCRYPTvol#1{\ifnum#1=
92{718}\fi}
\def\AUSCRYPTed#1{\ifnum#1=
92{}\fi}

\def\ASIACRYPTvol#1{\ifnum#1=
91{739}\else{\ifnum#1= 94{917}\else{\ifnum#1=
96{1163}\else{??}\fi}\fi}\fi}

\def\ASIACRYPTed#1{\ifnum#1=
91{H.~Imai, R.~Rivest and T.~Matsumoto}\else{\ifnum#1=
94{J.~Pieprzyk and R.~Safavi-Naini}\else{\ifnum#1= 96{M.~Y.~Rhee
and K.~Kim}\else{??}\fi}\fi}\fi}

\def\ICALPvol#1{\ifnum#1=
78{62}\else{\ifnum#1= 79{??}\else{\ifnum#1= 80{??}\else{\ifnum#1=
81{115}\else{\ifnum#1= 82{??}\else{\ifnum#1= 83{??}\else{\ifnum#1=
84{??}\else{\ifnum#1= 85{??}\else{\ifnum#1= 86{226}\else{\ifnum#1=
87{267}\else{\ifnum#1= 88{317}\else{\ifnum#1=
89{372}\else{\ifnum#1= 90{443}\else{\ifnum#1=
91{510}\else{\ifnum#1= 92{623}\else{\ifnum#1=
93{700}\else{\ifnum#1= 94{820}\else{\ifnum#1=
95{??}\else{\ifnum#1= 96{??}\else{\ifnum#1= 97{??}\else{\ifnum#1=
98{??}\else{\ifnum#1=
99{??}\fi}\fi}\fi}\fi}\fi}\fi}\fi}\fi}\fi}\fi}\fi}\fi}\fi}\fi}\fi}\fi}\fi}
\fi}\fi}\fi}\fi}\fi}

\def\ICALPed#1{\ifnum#1=
78{??}\else{\ifnum#1= 79{??}\else{\ifnum#1= 80{??}\else{\ifnum#1=
81{??}\else{\ifnum#1= 82{??}\else{\ifnum#1= 83{??}\else{\ifnum#1=
84{??}\else{\ifnum#1= 85{??}\else{\ifnum#1=
86{L.~Kott}\else{\ifnum#1= 87{T.~Ottman}\else{\ifnum#1=
88{T.~Lepisto}\else{\ifnum#1= 89{G.~Ausiello}\else{\ifnum#1=
90{M.~Paterson}\else{\ifnum#1= 91{A.~Leach}\else{\ifnum#1=
92{W.~Kuich}\else{\ifnum#1= 93{A.~Lingas}\else{\ifnum#1=
94{S.~Abiteboul}\else{\ifnum#1= 95{??}\else{\ifnum#1=
96{??}\else{\ifnum#1= 97{??}\else{\ifnum#1= 98{??}\else{\ifnum#1=
99{??}\fi}\fi}\fi}\fi}\fi}\fi}\fi}\fi}\fi}\fi}\fi}\fi}\fi}\fi}\fi}\fi}\fi}
\fi}\fi}\fi}\fi}\fi}

\def\svvconf#1#2{%
{\sl Proceedings of the\/} 19#2\ \csname #1name\endcsname, Lecture
Notes in Computer Science Vol.~\csname #1vol\endcsname{#2},
\csname #1ed\endcsname{#2} ed., Springer-Verlag, 19#2}

\def\FSTTCSvol#1{\ifnum#1=
94{880}\fi}
\def\FSTTCSed#1{\ifnum#1=
94{P.~Thiagarajan}\fi}

\def\STACSvol#1{\ifnum#1=
84{166}\else{\ifnum#1= 85{182}\else{\ifnum#1=
86{210}\else{\ifnum#1= 87{247}\else{\ifnum#1=
88{294}\else{\ifnum#1= 89{349}\else{\ifnum#1=
90{415}\else{\ifnum#1= 91{480}\else{\ifnum#1=
92{577}\else{\ifnum#1= 93{665}\else{\ifnum#1=
94{775}\else{\ifnum#1= 95{900}\else{\ifnum#1=
96{??}\else{\ifnum#1= 97{??}\else{\ifnum#1= 98{??}\else{\ifnum#1=
99{??}\fi}\fi}\fi}\fi}\fi}\fi}\fi}\fi}\fi}\fi}\fi}\fi}\fi}\fi}\fi}\fi}

\def\STACSed#1{\ifnum#1=
84{M.~Fontet}\else{\ifnum#1= 85{K.~Melhorn}\else{\ifnum#1=
86{B.~Monien}\else{\ifnum#1= 87{F.~Brandenburg}\else{\ifnum#1=
88{R.~Cori}\else{\ifnum#1= 89{B.~Monien}\else{\ifnum#1=
90{C.~Choffrut}\else{\ifnum#1= 91{C.~Choffrut}\else{\ifnum#1=
92{A.~Finkel}\else{\ifnum#1= 93{A.~Finkel}\else{\ifnum#1=
94{P.~Enjalbert}\else{\ifnum#1= 95{E.~Mayr}\else{\ifnum#1=
96{??}\else{\ifnum#1= 97{??}\else{\ifnum#1= 98{??}\else{\ifnum#1=
99{??}\fi}\fi}\fi}\fi}\fi}\fi}\fi}\fi}\fi}\fi}\fi}\fi}\fi}\fi}\fi}\fi}

\newcommand{\ccs}[1]{\ifnum#1=
93{{\sl Proceedings of the First Annual Conference on Computer and
Communications Security\/}, ACM, 1993}\else{\ifnum#1= 94{{\sl
Proceedings of the Second Annual Conference on Computer and
Communications Security\/}, ACM, 1994}\else{\ifnum#1=
96{{\sl Proceedings of the Third Annual Conference on Computer and
Communications Security\/}, ACM, 1996}\else{\ifnum#1= 97{{\sl
Proceedings of the Fourth Annual Conference on Computer and
Communications Security\/}, ACM, 1997}\else This CCS not yet
defined! \fi}\fi}\fi}\fi}

\newcommand{\dimacs}[1]{\ifnum#1=
89{{\sl Distributed Computing and Cryptography\/}, DIMACS Series
in Discrete Mathematics and Theoretical Computer Science, Vol.~2,
ACM, 1991}\else This DIMACS not yet defined! \fi}

\newcommand{\isaac}[1]{\ifnum#1=
92{{\sl Proceedings of ISAAC~92,\/} Lecture Notes in Computer
Science Vol.~650, Springer Verlag, 1992}\else This ISAAC not yet
defined! \fi}

\newcommand{\colt}[1]{\ifnum#1=
92{{\sl Proceedings of the Fifth Annual Workshop on Computational
Learning Theory\/}, ACM, 1992}\else This COLT not yet defined!
\fi}

\newtheorem{thm}{Theorem}
\newtheorem{lem}[thm]{Lemma}
\newtheorem{cor}[thm]{Corollary}
\newtheorem{propo}[thm]{Proposition}
\newtheorem{defn}[thm]{Definition}
\newtheorem{assm}[thm]{Assumption}
\newtheorem{clm}[thm]{Claim}
\newtheorem{rem}[thm]{Remark}
\newtheorem{examp}{Example}

\newenvironment{theorem}{\begin{thm}}
{
\end{thm}}
{
\end{lem}}
{
\end{cor}}
{
\end{propo}}
\newenvironment{definition}{\begin{defn}}
{
\end{defn}}
{
\end{assm}}
{
\end{clm}}
{\end{em}\end{rem}}
{\end{em}\end{examp}}

\newcount\proofqeded
\newcount\proofended
\def\qed{
\end{rm}\addtolength{\parskip}{-0pt}
\setlength{\parindent}{\saveparindent} \global\advance\proofqeded
by 1 }
\newenvironment{proof}%
 {\proofstart}%
 {\ifnum\proofqeded=\proofended\qed\fi \global\advance\proofended by 1
  \medskip}
\makeatletter
\def\proofstart{\@ifnextchar[{\@oprf}{\@nprf}}
\def\@oprf[#1]{\begin{rm}\protect\vspace{0pt}\noindent{\bf Proof of #1 } }
\def\@nprf{\begin{rm}\protect\vspace{0pt}\noindent{\bf Proof } }
\makeatother

\newcounter{ctr}
\newcounter{ectr}
\newcounter{eectr}

\newlength{\savejot}
\newenvironment{newmath}{\begin{displaymath}%
\setlength{\abovedisplayskip}{5pt}%
\setlength{\belowdisplayskip}{5pt}%
\setlength{\abovedisplayshortskip}{3pt}%
\setlength{\belowdisplayshortskip}{3pt} }{\end{displaymath}}

\def\poly{\mathop{\rm poly}\nolimits}
\newcommand{\eqdef}{\stackrel{\rm def}{=}}

\def\e{\epsilon}

\newcommand{\bits}{\{0,1\}}

\newcommand{\xor}{{\oplus}}

\newcommand{\Colon}{{:\;\;}}

\def\leqq{{\:\leq\:}}

\newcommand{\Z}{{\hbox{\sf Z}}}
\newcommand{\R}{{\hbox{\rm\bf R}}}
\newcommand{\N}{{\hbox{\bf N}}}

\def\next{\: ; \:}

\newcommand{\ProbExp}[2]{{{\Pr}\left[{#1}\: : \:#2\right]}}

\def\next{\:;\:}

\def\e{\epsilon}

\begin{document}

\maketitle

\thispagestyle{first} \setcounter{page}{245}

\begin{abstract}

\vskip 3mm

Theoretical computer science has found fertile ground in many
areas of mathematics. The approach has been to consider classical
problems through the prism of computational complexity, where the
number of basic computational steps taken to solve a problem is
the crucial qualitative parameter.  This new approach has led to a
sequence of advances, in setting and solving new mathematical
challenges as well as in harnessing discrete mathematics to the
task of solving real-world problems.

In this talk, I will survey the development of modern cryptography
--- the mathematics behind secret communications and protocols ---
in this light.  I will describe the complexity theoretic
foundations underlying the cryptographic tasks of encryption,
pseudo-randomness number generators and functions, zero knowledge
interactive proofs, and multi-party secure protocols. I will
attempt to highlight the paradigms and proof techniques which
unify these foundations, and which have made their way into the
mainstream of complexity theory.

\vskip 4.5mm


\end{abstract}

\vskip 12mm

\section{Introduction} \label{section 1} \setzero
\vskip-5mm \hspace{5mm}

The mathematics of cryptography is driven by real world
applications. The original and most basic application is the wish
to communicate privately in the presence of an eavesdropper who is
listening in. With the rise of computers as means of
communication, abundant other application arise, ranging from
verifying authenticity of data and access priveleges to enabling
complex financial transactions over the internet involving several
parties each with its own confidential information.

As a rule, in theoretical fields inspired by applications, there
is always a subtle (and sometimes not so subtle) tension between
those who do ``theory'' and those who ``practice''. At times, the
practitioner shruggs of the search for a provably good method,
saying that in practice his method works and will perform much
better when put to the test than anything for which a theorem
could be proved. The theory of Cryptography is unusual in this
respect. Without theorems that provably guarantee the security of
a system, it is in a sense worthless, as there is no observable
outcome of using a security system other than the guarantee that
no one will be able to crack it.



In computational complexity based cryptography one takes feasible
(or easy) to mean those computations that terminate in polynomial
time and infeasible (or hard) those computations that do
not\footnote{ We remark however that all security definitions
(although not necessarily all security proofs) still make sense
for a different meaning of `easy' and `hard'. For example, one may
take easy to mean linear time whereas hard to mean quadratic time.
)}. Achieving many tasks of cryptography relies on a gap between
feasible algorithms used by the legitimate user versus the
infeasibility faced by the adversary. On close examination then,
it becomes apparent that a necessary condition for many modern
cryptographic goals is that $NP\neq P$ \footnote{This is the
celebrated unresolved NP vs. P problem posed by Karp, Cook and
Levin in the early seventies. NP corresponds to those problems for
which given a solution its correctness be verified in polynomial
time whereas P corresponds to those problems for which a solution
can be found in polynomial time.}, although it is not known to be
a sufficient condition. A (likely) stronger necessary condition
which is also sufficient for many tasks is the existence of {\it
one-way functions}:  those functions which are easy to compute but
hard to invert with non-negligible probability of success taken
over a polynomial time samplable distribution of inputs.

In 1976 when Diffie and Hellman came out with their paper ``New
Direction in Cryptography'' \cite{DH76} announcing that we are
``on the brink of a revolution in cryptoghraphy'' hopes were high
that the resolution of the celebrate $P$ vs. $NP$ problem was
close at hand and with it techniques to lower bound the number of
steps required to break cryptosystems. That did not turn out to be
the case. As of today,  no non-linear lower bounds are known for
any $NP$ complete problem\footnote{ NP-complete problems are the
hardest problems for $NP$. Namely, if an NP complete problem can
be solved in polynomial time and thus be in P, then all problems
in NP are in P.}.

Instead, we follow a 2-step program when faced with a
cryptographic task which can not be proved unconditionally (1)
find the minimal assumptions necessary and sufficient for the task
at hand. (2) design a cryptographic system for the task and prove
its security if and only if the minimal assumptions hold. Proofs
of security then are realy proofs of secure design. They take a
form of a constructive reduction. For example, the existence of a
one-way function has been shown a sufficient and necessary
condition for ``secure'' digital signatures to
exist\cite{GMRi88,NY89, R90}. To prove this statement one must
show how to convert any ``break'' of the digital signature scheme
into an efficient algorithm to invert the underlying one-way
function. Defining formally ``secure'' and ``break'' is an
essential preliminary step in accomplishing this program.

These type of constructive reductions are a double edged sword.
Say that system has been proved secure if and only if integer
factorization is not in polynomial time. Then, either the system
is breakable and then the reduction proof immediately yields a
polynomial time integer factorization algorithm which will please
the mathematicians to no end, or there exists no polynomial time
integer factorization algorithms and we have found a superb
cryptosystem with guaranteed security which will please the
computer users to no end.


Curiously, whereas early hopes of complexity theory producing
lower bounds have not materialized, cryptographic research has
yielded many dividends to complexity theory. New research themes
and paradigms, as well as techniques originating in cryptography,
have made their way to the main stream of complexity theory. Well
known techniques include random self-reducibility, hardness
amplification, low degree polynomial representations of Boolean
functions, and proofs by hybrid and simulation arguments. Well
known examples of research themes include : interactive and
probabilisticly checkable proofs and their application to show
inapproximability of NP-hard algorithmic problems, the study of
average versus worst case hardness of functions, and trading off
hardness of computation for randomness to be used for
derandomizing probabilistic complexity classes.

These examples seem, on a superficial level, quite different from
each other. There are similarities however, in addition to the
fact that they are investigated by a common community of
researchers, who use a common collection of techniques. In all of
the above, an ``observer'' is always present, success and failure
are defined ``relative to the observer'', and if the observer
cannot ``distinguish'' between two probabilistic events, they are
treated as identical. This is best illustrated by examples. (1) A
probabilistically checkable proofs is defined to achieve soundness
if the process of checking it errs with exponentially small
probability (which is indistinguishable from zero). (2) A function
is considered hard to compute if all observers fail to compute it
with non negligible probability taken over a efficiently samplable
input distribution. It is not considered ``hard'' enough if it is
only hard to compute with respect to some worst case input never
to be encountered by the observer. (3) A source outputting bits
according to some distribution is defined as pseudorandom if no
observer can distinguish it from a truly random  source
(informally viewed as an on going process of flipping a fair
coin).

\subsection{Cryptography and classical mathematics}

\vskip-5mm \hspace{5mm}

Computational infeasibility, which by algorithmic standards is the
enemy of progress, is actually the cryptographer's best friend.
When a computationally difficult problem comes along with some
additional properties to be elaborated on in this article, it
allows us to design methods which while achieving their intended
functionality are ``infeasible'' to break. Luckily, such
computationally intensive problems are abundant in mathematics.
Famous examples include {\it integer factorization, finding short
vectors in an integer lattice, and elliptic curve logarithm
problem}. Viewed this way, cryptography is an external customer of
number theory, algebra, and geometry. However, the complexity
theory view point has not left these fields untouched, and often
shed new light on old problems.

In particular, the history of cryptography and complexity theory
is intertwined with the development of algorithmic number theory.
This is most evident in the invention of faster tests for integer
primality testing and integer factorization \cite{LL90} whose
quality is attested by complexity analysis rather than the earlier
benchmarking of their performance. A beautiful account on the
symbiotic relationship between number theory and complexity theory
is given by Adleman \cite{Adleman94} who prefaces his article by
saying that ``Though algorithmic number theory is one of man's
oldest intellectual pursuit, its current vitality is unrivaled in
history. This is due in part to the injection of new ideas from
computational complexity.''

\subsection{Cryptography and information theory}

\vskip-5mm \hspace{5mm}

In a companion paper to his famous paper on information theory,
Shannon \cite{Shannon49} introduced a rigorous theory of perfect
secrecy based on information theory. The theory addresses
adversary algorithms which have unlimited computational resources.
Thus, all definitions of security, which we will refer to
henceforth as {\it information theoretic security}, and proofs of
possibility and impossibility are with respect to such adversary.
Shannon proves that ``perfectly secure encryption'' can only exist
if the size of secret information that legitimate parties exchange
between them in person prior to remote transmission, is as large
as the total entropy of secret messages they exchange remotely.
Maurer \cite{Mau93} generalized these bounds to two-way
communications. This limits the practice of encryption based on
information theory a great deal. Even worse, the modern
cryptographic tasks of public-key encryption, digital signatures,
pseudo random number generation, and most two party protocols can
be proved down right impossible information theoretically. To
achieve those, we turn to adversaries who are limited
computationally and aim at computational security with the cost of
making computational assumptions or assumptions about the physical
world.

Having said that, some cryptographic tasks can achieve full
information theoretic security. A stellar example is of multi
party computation. Efficient and information theoretic secure
multi-party protocols are possible unconditionally tolerating less
than half faults, if there are perfect private channels between
each pair of honest users \cite{BGW88,CCD88,RB89,GL02}.
Statistical {\it zero-knowledge proofs} are another example
\cite{GMR86,SA}.

Perfect private channels between pairs of honest users can be
implemented in several settings: (1) The {\it noisy channel}
setting \cite{Kilian88} (which is a generalization of the {\it
wire tal channel} \cite{W75}) where the communication between
users in the protocol as well as what the adversary taps is
subject to noise). (2) A setting where the adversary's memory
(i.e. ability to store data) is limited \cite{MC97}. (3) The {\it
Quantum Channels} setting where by quantum mechanics, it is
impossible for the adversary to obtain full information on
messages exchanged between honest users. Introducing new and
reasonable such settings which enable information theoretic
security is an important activity.

Moreover, often paradigms and construction introduced within the
computational security framework can be and have been lifted out
to achieve information theoretic security. The development of
randomness extractors from pseudo random number generators can be
done in this fashion \cite{Tr}.

We note that whereas the computational complexity notions of
secrecy, knowledge, and pseudo-randomness are different than their
information theoretic analogues, techniques of error recovery
developed in information theory are extremely useful. Examples
include the Haddamard error correcting codes which is used to
exhibit hard core predicates in one-way functions \cite{GL}, and
various polynomial based error correcting codes which enable high
fault tolerance in multi-party computation \cite{BGW88}.

To sum up, the theory of cryptography has in the last 30 years
turned into a rich field with its own rules, structure, and
mathematical beauty which has helped to shape complexity theory.
In the talk, I will attempt to lead you through  a short summary
of  what I believe to have been a fascinating journey of modern
cryptography. I apologize in advance for describing my own
journey, at the expense of other points of view. I attach a list
of references including several survey articles that contain full
details and proofs \cite{book}.

In the rest of the article, I will briefly reflect on a few points
which will make my lecture easier to follow.

\section{Conventions and complexity theory terminology}

\vskip-5mm \hspace{5mm}

We say that an algorithm is {\it polynomial time} if for all
inputs $x$, the algorithm runs in time bounded by some polynomial
in $|x|$ where the latter denotes the length of $x$ when
represented as a binary string. A {\it probabilistic algorithm} is
one that can make random choices, where without loss of generality
each choice is among two and is taken with probability 1/2. We
view these choices as the algorithm {\it coin tosses}. A
probabilistic algorithm $A$ on input $x$ may have more than one
possible output depending on the outcome of its coin tosses, and
we will let $A(x)$ denote the probability distribution over all
possible outputs. We say that a probabilistic algorithm is {\it
probabilistic polynomial time} (PPT) if for any input $x$, the
expectation of the running time taken over the all possible coin
tosses is bounded by some polynomial in $|x|$, regardless of the
outcome of the coin tosses.

In complexity theory, we often speak of language classes. A language is
a subset of all binary strings. The class P is the set of languages such
that there exists a polynomial time algorithm, which on every input $x$
can decide if $x$ is in the language or not. The class ${\rm BPP}$ are
those languages whose membership can be decided by a probabilistic
polynomial time algorithm which for every input, is incorrect with at
most negligible probability taken over the coin tosses of the algorithm.
The class NP is the class of languages accepted by polynomial time
non-deterministic algorithm which may make non-deterministic choices at
every point of computation. Another characterization of NP is as the
class of languages that have short proofs of memberships.  Formally,
$NP=\left\{ L|\right.$ there exists polynomial time computable function
$f$ and $k >$ 0, such that $x\in L$ iff there exists $y$ such that
$f(x,y)= 1$ and $\left.|y|<|x|^{k}\right\}$.

In this article, we consider an `easy' computation to be one which
is carried out by a PPT algorithm. A function $\nu\Colon\N\to\R$
is negligible if it vanishes faster than the inverse of any
polynomial. All probabilities are defined with respect to finite
probability spaces.

\section{Indistinguishability}

\vskip-5mm \hspace{5mm}

Indistinguishability of probability distributions is a central
concept in modern cryptography. It was first introduced in the
context of defining security of encryption systems by Goldwasser
and Micali \cite{GM82}. Subsequently, it turned out to play a
fundamental role in defining pseudo-randomness by Yao \cite{Y82},
and zero-knowledge proofs by Goldwasser, Micali, and Rackoff
\cite{GMR86}.

\begin{defn}
Let $X=\{X_{k}\}_k$, $Y=\{Y_{k}\}$ be two ensembles of probability
distributions on $\{0,1\}^{k}$. We say that $X$ is {\bf
computationally indistinguishable} from $Y$ if $\forall$
probabilistic polynomial time algorithms  $A$, $\forall$ $c>0$,
$\exists k_{0}$, s.t $\forall k > k_{0}$,
\[ |\Pr_{t \in X_{k}}(A(t) = 1) - \Pr_{t \in Y_{k}}(A(t) = 1) | <
\frac{1}{k^c}. \] The algorithm $A$ used in the above definition
is called a polynomial time {\em statistical test}.
\end{defn}

Namely, for sufficiently long strings, no probabilistic polynomial
time algorithms can tell whether the string was sampled according
to $X$ or according to $Y$.  Note that such a definition cannot
make sense for a single string, as it can be drawn from either
distribution. Although we chose to focus on polynomial time
indistinguishability, one could instead talk of distribution which
are indistinguishable with respect to any other computational
resource, in which case all the algorithms $A$ in the definition
should be bounded by the relevant computational resource. This,
has been quite useful when applied to space bounded computations
\cite{Ni}.

Of particular interest are those probability distributions which
are indistinguishable from the uniform distribution, focused on in
\cite{Y82}, and are called {\it pseudorandom distributions}.

Let $U=\{U_k\}$ denote the uniform probability distribution on
$\{0,1\}^k$.  That is, for every $\alpha \in\{0,1\}^k$, $\Pr_{x
\in U_k} [x = \alpha] = \frac{1}{2^k}$.

\begin{defn}
We say that $X=\{X_k \}_k$ is {\bf pseudo random} if it is
computationally indistinguishable from $U$. That is, $\forall$
probabilistic polynomial time algorithms  $A$, $\forall$ $c>0$
$\exists k_{0}$, such that $\forall k > k_{0}$,
\[ |\Pr_{t \in X_k}[A(t) = 1] - \Pr_{t \in U_k}[A(t) = 1]| <
\frac{1}{k^c}. \] If $\exists A$ and $c$ such that the condition
in definition $2$ is violated, we say that $X_k$ fails the
statistical test $A$.
\end{defn}

A simple but not very interesting example of two probability
distributions which are computationally indistinguishable are two
distributions which are statistically very close. For example,
$X=\{X_k\}$ defined exactly as the uniform distribution over
$\{0,1\}^k$ with two exceptions, $0^k$ appears with probability
${1\over{2^{k+1}}}$ and $1^k$ appears with probability
${3\over{2^{k+1}}}$. Then the uniform distribution and $X$ can not
be distinguished by any algorithm (even one with no computational
restrictions) as long as it is only given a polynomial size sample
from one of the two distributions.

It is fair to ask as this point whether computationally
indistinguishability is anything more than statistical closeness
where the latter is formally defined as follows.
\begin{defn}
Two probability distributions $X,Y$ are statistically close if
$\forall c>0$, $\exists k_0$ such that $\forall k>k_0$,
\[ \sum_t|\Pr({t \in X_k})- \sum_t({t \in U_k})< \frac{1}{k^c}. \]
$X$ and $Y$ are {\it far} if they are not close.
\end{defn}

Do there exist distributions which are statistically far apart and
yet are computationally indistinguishable? Goldreich and Krawczyk
\cite{GK89} who pose the question note this to be the case by a
counting argument. However their argument is non constructive. The
works on secure encryption and pseudo random number generators
\cite{GM82, BM82, Y82} imply the existence of {\it efficiently
constructible} pairs of distributions that are computationally
indistinguishable but statistically far, under the existence of
one-way functions. The use of assumptions is no accident.

\begin{thm}{\rm \cite{Go}}
The existence of one-way functions is equivalent to the existence
of pairs of polynomial-time constructible distributions which are
computationally indistinguishable  and statistically far.
\end{thm}

\section{Building blocks}

\vskip-5mm \hspace{5mm}

A central building block required for many tasks in cryptography
is the existence of a one-way function. Let us discuss this basic
primitive as well as a few others in some detail.

\subsection{One-way functions}\label{ow-def.sec}

\vskip-5mm \hspace{5mm}

Informally, a one-way function is a function which is ``easy'' to
compute but ``hard'' to invert. Any probabilistic polynomial time
(PPT) algorithm attempting to invert the function on an element in
its range, should succeed with no more than ``negligible''
probability, where the probability is taken over the elements in
the domain of the function and the coin tosses of the PPT
attempting the inversion. We often refer to an algorithm
attempting to invert the function as an adversary algorithm.

\begin{defn}
A function $f\Colon\bits^*\to\bits^*$ is {\em one-way\/} if:
\begin{enumerate}
\item Easy to Evaluate: there exists a PPT algorithm that on input $x$ output $f(x)$;
\item Hard to Invert: for all PPT algorithm $A$, for all $c>0$,
there exists $k_0$ such that for all $k>k_0$,
\begin{newmath}
{ \ProbExp{A(1^k,f(x)) =z}{f(x)=f(z)}} \leqq {1\over{k^c}}
\end{newmath}%
where the probability is taken over $x\in \{0,1\}^k$ and the coin
tosses of $A$.
\end{enumerate}
\end{defn}

\noindent{\bf Note} Unless otherwise mentioned, the probabilities
during this section are calculated uniformly over all coin tosses
made by the algorithm in question.

A few remarks are in order. (1)The guarantee is probabilistic. The
adversary has low probability of inverting the function where the
probability distribution is taken over the inputs of length $k$ to
the one-way function and the possible coin tosses of the
adversary.

(2) The adversary is not asked to find $x$; that would be pretty
near impossible. It is asked to find some inverse of $f(x)$.
Naturally, if the function is 1-1 then the only inverse is $x$. We
note that it is much easier to find candidate one-way functions
without imposing further restrictions on its structure, but being
1-1 or at least {\it regular} (that is, the number of preimage of
any image is about of the range), it results in easier and more
efficient cryptographic constructions.

(3) One may consider a non-uniform version of the ``Hard to
invert'' requirement, requiring the function to be hard to invert
by all non-uniform polynomial size family of algorithms, rather
than by all probabilistic polynomial time algorithms. The former
extends probabilistic polynomial time algorithms to allow for each
different input size, a different polynomial size algorithm.

(4) The definition is typical to definitions from computational
complexity theory, which work with asymptotic complexity---what
happens as the size of the problem becomes large. One-wayness is
only asked to hold for large enough input lengths, as $k$ goes to
infinity. Per this definition, it may be entirely feasible to
invert $f$ on, say, 512 bit inputs. Thus such definitions are
useful for studying things on a basic level, but need to be
adapted to be directly relevant to practice.

(5) The above definition can be considerably weakened by replacing the
second requirement of the function to require it to be hard to invert on
{\bf some} non-negligible fraction of its inputs (rather than all but
non-negligible fraction of its inputs ). This relaxation to a {\it weak
one-way function} is motivated by the following example. Consider the
function $f:{\bf Z} \times {\bf Z} \mapsto {\bf Z}$ where $f(x,y) = x
\cdot y$.  This function can be easily inverted on at least half of its
outputs (namely, on the even integers) and thus is not a one-way
function as defined above. Still, $f$ resists all efficient algorithms
when $x$ and $y$ are primes of roughly the same length which is the case
for a non-negligible fraction ($\approx {1\over{{k^2}}}$) of the $k$-bit
composite integers. Thus according to our current state of knowledge of
integer factorization, $f$ does satisfy the weaker requirement.
Convertion between any weak one-way function to a one-way function have
been shown using ``hardness amplification'' techniques which expand the
size of the input by a polynomial factor \cite{Y82}. Using expanders,
constant factor expansions (of the input size) construction of a one-way
function from a weak one-way function is possible \cite{GILVZ}.

(6) To apply this definition to practice we must typically
envisage not a single one-way function but a family of them,
parameterized by a {\em security parameter\/} $k$.  That is, for
each value of the security parameter $k$, there is a family of
functions, each defined over some finite domain and finite ranges.
The existence of a single one-way function is equivalent to the
existence of a collection of one-way functions.

\begin{defn}\label{oneway}
A collection of one-way functions is a set $F = \{ f_i: D_i
\rightarrow R_i \}_{i \in I}$ where $I$ is an index set, and $D_i$
($R_i$) are finite domain(range) for $i\in I$, satisfying the
following conditions.
\begin{enumerate}
\item Selection in Collection:
$\exists$ PPT algorithm $S_1$ that on input $1^k$ outputs an $i\in
I$ where $|i|=k$.
\item
Selection in Domain: $\exists$ PPT algorithm $S_2$ that on input
$i \in I$ outputs $x \in D_i$
\item
Easy to Evaluate: $\exists$ PPT algorithm $Eval$ such that for $i
\in I$ and $x \in D_i$, $Eval(i,x) = f_i(x)$.
\item
Hardness to Invert: $\forall$ PPT adversary algorithm $A$, $c>0$,
$\exists$ $k_0$ such that $\forall$ $k>k_0$,
 $${\ProbExp{A(1^k,i, f_i(x)) =z}{f(x)=f(z)}} \leqq {1\over{k^c}}$$
(the probability is taken over $i\in S_1(1^k),x\in S_2(i)$ and the
coin tosses of $A$).
\end{enumerate}
\end{defn}

The hardness to invert condition can be made weaker by requiring
only that $\exists c >0$, such that $\forall$ PPT algorithm $A$,
$\exists$ $k_0$ such that $\forall$ $k>k_0$. $Prob[A(1^k,i,
f_i(x)) \neq z$, ${f(x)=f(z)}] > {1\over{k^c}}$ (the probability
taken over $i\in S_1(1^k),x\in S_2(i)$ and the coin tosses of
$A$). We call collections which satisfy such weaker conditions,
collection of weak one-way functions. Transformations exist via
sampling algorithms between both types of collections.

Another useful and equivalent notion is of a {\em one-way
predicate}, first introduced in \cite{GM82}. This is a Boolean
function of great use in encryption and protocol design. A one-way
predicate is equivalent to the existence of 0/1 problems, for
which it is possible to uniformly select an instance for which the
answer is 0 (or respectively 1), and yet for a (pre-selected)
instance it is hard to compute with success probability greater
than $1\over 2$ whether the answer is 0 or 1.

\begin{defn}A {\it one-way predicate }
is a Boolean function $B:\{0,1\}^*\rightarrow\{0,1\}$ for which
\begin{enumerate}
\item {\it Sampling is possible}:
$\exists$ PPT algorithm $S$ that on input $v\in\{0,1\}$ and $1^k$,
outputs a random $x$ such that $B(x)= v$ and $x\in\{0,1\}^k$.
\item {\it Guessing is hard}:
$\forall c>0$, $\forall$ PPT algorithms $A$,  $\forall k$
sufficiently large, $Prob[A(x)$ $=B(x)] \leqq {{1\over 2
}+{1\over{k^c}}}$ (probability is taken over $v\in \{0,1\}, x \in
S(1^k,v)$, and the coin tosses of $A$).
\end{enumerate}
\end{defn}

Proving the equivalence between one-way predicates and one-way
functions is easy in the forward direction, by viewing the
sampling algorithm $S$ as a function over its coin tosses. To
prove the reverse implication is quite involved. Toward this goal,
the notion of a hard core predicate of a one-way function was
introduced in \cite{BM82,Y82}. Jumping ahead, hard core predicate
of one-way functions yield immediately one-way predicates.

\subsubsection{Hard-core predicates}
\label{Hard-Core}

\vskip-5mm \hspace{5mm}

The fact that $f$ is a one-way function obviously  does not
necessarily imply that $f(x)$ hides everything about $x$. It is
easy to come up with constructions of universal one-way functions
in which one of the bits of $x$ {\it leaks} from $f(x)$. Even if
each bit of $x$ is well hidden by $f(x)$ then some function of all
of the bits of $x$ can be easy to compute. For example, the least
significant bit of $x$ is easy to compute from $f_{p,g}(x) =
g^x\bmod p$ where $p$ is a prime and $g$ a generator for the
cyclic group $Z_p^*$, even though we know of no polynomial time
algorithms to compute $x$ from $f_{p,g}(x)$.  Similarily, it is
easy of compute the Jacobi symbol of $x$ mod $n$ from the RSA
function $RSA_{n,e}(x)=x^e\bmod n$ where $(e,\phi(n))=1$, even
though the fastest algorithm to invert $RSA_{n,e}$ needs to factor
integer $n$ first, which is not known to be a polynomial time
computation.

Yet, clearly there are some bits of information about $x$ which
cannot be computed from $f(x)$, given that $x$ in its entirety is
hard to compute. The question is, which bits of $x$ are hard to
compute, and how hard to compute are they. The answer is
encouraging. For several functions $f$ for which no polynomial
time inverting algorithm is known, we can identify  particular
bits of the pre-image of $f$ which can be proven (via a polynomial
time reduction) to be as hard as to compute with probability
significantly better than $1\over 2$, as it is to invert $f$
itself in polynomial time. Examples of these can be found in
\cite{BM82,GM82,GMT,ACGS88}.

More generally, a {\it hard-core predicate} for $f$, is a Boolean
predicate about $x$ which is efficiently computable given $x$, but
is hard to compute from $f(x)$ with probability significantly
better than $1\over 2$.

\begin{defn}
A hard-core predicate of a function $f:\{0,1\}^* \rightarrow
\{0,1\}^*$ is a Boolean predicate $B:\{0,1\}^* \rightarrow
\{0,1\}$, such that
\begin{enumerate}
\item $\exists PPT$ algorithm $Eval,$ such that $\forall x \, Eval(x)=B(x)$
\item $ \forall \,\mbox{PPT algorithm} \,A,\ \forall \,c>0,\ \exists k_0,\ \mbox{s.t.}\, \forall {k>k_0}$
 $\Pr[A(f(x)) = B(x)] < \frac{1}{2}+\frac{1}{k^c}.$
The probability is taken over the random coin tosses of A,  and
random choices of $x$ of length $k$.
\end{enumerate}
\end{defn}

Yao proposed a construction of a hard-core predicate for any
one-way function \cite{Y82}.  A  considerably simpler construction
and proof general result is due to Goldreich and Levin \cite{GL}.

\begin{thm}\label{th-gl} {\rm \cite{GL}}
Let $f$ be a length preserving one-way function. Define $f'(x
\circ r)=f(x) \circ r$, where $|x|=|r|=k$, and $\circ$ is the
concatenation function. Then
\[ B(x \circ r) = {\displaystyle \Sigma_{i=1}^{k}} x_ir_i (mod\ 2)\]
is a hard-core predicate for $f'$ (Notice that if $f$ is one-way
then so is $f'$).
\end{thm}

\medskip

Interestingly, the proof of the theorem can be regarded as the
first example of a polynomial time list decoding \cite{Su}
algorithm. Essentially $B(x,r)$ may be viewed as the $r$th bit of
a Haddamrd encoding of $x$. The proof of the theorem yields a
polynomial time error decoding algorithm which returns a
polynomial size list of candidates for $x$, as long as the
encoding is subject to an error rate of less than ${1\over 2} -
{\epsilon}$ where $\epsilon > {1\over{k^c}}$ for some constant
$c>0$, $k=|x|$. The length of the list is
$O({1\over{\epsilon^2}})$.

\subsection{Trapdoor functions }\label{sec-trapdoor}

\vskip-5mm \hspace{5mm}

A {\em trapdoor function\/} $f$ is a one-way function with an
extra property.  There also exists a secret inverse function (the
{\em trapdoor\/}) that allows its possessor to efficiently invert
$f$ at any point in the domain of his choosing.  It should be easy
to compute $f$ on any point, but infeasible to invert $f$ with
high probability without knowledge of the inverse function.
Moreover, it should be easy to generate matched pairs of $f$'s and
corresponding trapdoor.

\begin{defn}
A {\it trapdoor} function is a one-way function
$f:\{0,1\}^*\rightarrow \{0,1\}^*$ such that there exists a
polynomial $p$ and a probabilistic polynomial time algorithm $I$
such that for every $k$ there exists a $t_k\in\{0,1\}^*$ such that
$|t_k|\leq p(k)$ and  for all $x\in \{0,1\}^k$, $I(f(x), t_k) =y$
such that $f(y)=f(x)$.
\end{defn}

%

Trapdoor functions are much harder to locate than one-way
function, as they seem to require much more hidden structure. An
important problem is to establish whether one implies the other.
Recent results of \cite{RMG} indicate this may not the case.

A {\em trapdoor predicate\/} is a one-way predicate with an extra
trapdoor property: for every $k$, there must exist trapdoor
information $t_k$ whose size is bounded by a polynomial in $k$ and
whose knowledge enables the polynomial-time computation of $B(x)$,
for all $x\in\{0,1\}^k$. Restating as a collection of trapdoor
predicates we get.

\begin{defn}
Let $I$ be an index set and for $i \in I$,  $D_i$ a finite domain.
A collection of trapdoor is a set $B = \{ B_i: D_i \rightarrow
\{0,1\}\}_{i \in I}$ such that:
\begin{enumerate}
\item $\exists$ PPT algorithm $S_1$ which on input $1^k$
outputs $(i,t_i)$ where $i \in I \cap \{0,1\}^k$, and
$|t_i|<poly(k)$  ( $t_i$ is the trapdoor).
\item $\exists$ PPT algorithm $S_2$ which on input $i \in I, v\in \{0,1\}$
outputs $x \in D_i$ such that $B_i(x)=v$.
\item $\exists$ PPT algorithm $S_3$ which
on input $i\in I, x\in D_i, t_i$ outputs $B_i(x)$.
\item $\forall$ PPT adversary algorithms
$A$, $c>0$, $\exists k_0, \forall k> k_0$, $Prob[A(i,x)= B_i(x) ]
\leq {{1\over2} +{1\over{k^c}}}$ (the probability taken over $i\in
S_1({1^k}), v\in \{0,1\}, x\in S_2(i,v)$, and the coins of A).
\end{enumerate}
\label{one-way-predicates}
\end{defn}

The existence of a trapdoor predicate is equivalent to the
existence of secure public-key encryption as we shall see in the
next section. Trapdoor functions imply trapdoor predicates, but it
is an open problem to show that they are equivalent.

\begin{clm}
If trapdoor functions exist then collection of trapdoor predicates
exist.
\end{clm}

\subsection{Candidate examples of building blocks}

\vskip-5mm \hspace{5mm}

It has been shown by a fairly straightforward diagonalization
argument \cite{GO1} how to construct a {\it universal} one-way
function (i.e. a function which is one-way if any one-way function
exists). Still this is very inefficient, and concrete proposals
for one-way function are needed for any practical usage of
cryptographic constructions which utilized one-way functions.
Moreover, looking into the algebraic, combinatorial, and geometric
structure of concrete proposals has lead to many insights about
what could be true about general one-way functions.  The
revelation process seems almost always to start from proving
properties about concrete examples to generalizing to proving
properties on general one-way functions.

Interesting proposals for one-way functions, trapdoor functions,
and trapdoor predicates have been based on hard computational
problems from number theory, coding theory, algebraic geometry,
and geometry of numbers. What makes a computational problem a
``suitable'' candidate? First, it should be put under extensive
scrutiny by the relevant mathematical community. Second, the
problem should be hard on the {\it average} and not  only in the
{\it worst} case. A big project in cryptography is the
construction of cryptographic functions which are provably hard to
break {\it on the average} under some {\it worst-case}
computational complexity assumption. A central technique is to
show that a problem is as hard for an {\it average} instance as it
is for a {\it worst case} instance by {\it random self
reducibility} \cite{AL}. A problem $P$ is random self reducible if
there exists a probabilistic polynomial time algorithm that maps
any instance $I$ of $P$ to a collection of random instances of $P$
such that given solutions to the random instances, one can
efficiently obtain a solution to the original instance. Variations
would allow mapping any instance of $P$ to random instances of
$P'$.\footnote{This technique was first observed and applied to
the number theoretic problems of factoring, discrete log, testing
quadratic residuosity, and the RSA function. In each of these
problems, one could use the algebraic structure to show how to map
a particular input uniformly and randomly to other inputs in such
a way that the answer for the original input can be recovered from
the answers for the targets of the random mapping. Showing that
polynomials are randomly self reducible over finite fields was
applied to the low-degree polynomial representations of Boolean
functions, and has been a central and useful technique in
probabilistically checkable proofs.}

Perhaps the most interesting problem in cryptography today is to
show (or rule out) that the existence of a one-way function is
equivalent to the $NP\neq BPP$.

For lack of space, we discuss in brief a few proposals.

\subsubsection{Discrete logarithm problem proposal}

\vskip-5mm \hspace{5mm}

Let $p$ be a prime integer and $g$ a generator for the
multiplicative cyclic group ${\cal Z}^{*}_{p} = \{ 1\leq y<p |
(y,p) = 1\}$. The {\it discrete log problem} (DLP) is given
$p$,$g$, and $y\in Z_p^*$, compute the unique $x$ such that $1\leq
x\leq p-1$ and $y=g^x\bmod p$. The discrete log problem has been
first suggested to be useful for key exchange over the public
channel by Diffie and Hellman \cite{DH76}.

The function $DL(p,g,x)=(p,g,g^x\bmod p)$, and the corresponding
collection of functions
$DL=\{DL_{p,g}:{Z_{p-1}}\rightarrow{Z_p^*}, DL_{p,g}(x)=g^x\bmod
p\}_{<p,g>\in I}$ where $I=\{<p,g>, p\hbox{ prime }, g\hbox{
generator}\}$ have served as proposals for a one-way function and
a collection of one-way functions (respectively). On one hand,
there exist efficient algorithms to select pairs of $(p,g)$ of a
given length with uniform probability \cite{Bach88}, and to
perform modulo exponentiation. On the other hand, the fastest
algorithms to solve the discrete log problem is the generalized
number field sieve version of the index-calculus method which runs
in expected time $e^{((c + o(1))(\log p)^{\frac{1}{3}}(\log \log
p)^{\frac{2}{3}})}$ (see survey \cite{odlyzko}). Moreover, for a
fixed prime $p$,  $DL(p,g,g^x\bmod p)$ can be shown as hard to
invert on the average over the $1\leq x\leq {p-1}$ and $g$
generators, as it is for every $g$ and $x$.

An important open problem is to prove that, {\it without fixing
first the prime $p$}, solving the discrete log problem for an
average instance $(p,g,y)$ is hard on the average as in the worst
case.

In the mid-eighties an extension of the discrete logarithm problem
over prime integers, to {\it computing discrete logarithms over
elliptic curves} was suggested by Koblitz and V. Miller (see
survey \cite{koblitz}). The attraction is that the fastest
algorithms known for computing logarithms over elliptic curves are
of complexity $O(\sqrt p )$ for finite field $F_p$. The main
concern is that they have not been around long enough to go under
extensive scrutiny, and that the intersection between the
mathematical community who can offer such scrutiny and the
cryptographic community is not large.

\subsubsection{Shortest vector in integer lattices proposal}

\vskip-5mm \hspace{5mm}

In a celebrated paper \cite{Aj} Ajtai described a problem that is
hard {\it on the average} if some well-known integer lattice
problems are {\em hard to approximate in the worst case}, and
demonstrated how this problem can be used to construct one-way
functions. Previous worst case to average case reductions were
applied to two parameter problems and the reduction was shown upon
fixing one parameter (e.g. in the discrete logarithm problem
random self reducibility was shown fixing the prime parameter),
whereas the \cite{Aj} reduction is the first which averages over
all parameters.

Let $V$ be a set of $n$ linearly independent vectors $V = \{v_1,
\cdots, v_n, v_i\in {\cal R}\}$. The integer lattice spanned by
$V$ is the set of all possible linear combinations of the $v_i$'s
with integer coefficients, namely $L(V) \eqdef \left\{\sum_i a_i
v_i\ :\ a_i\in{\Z}\mbox{ for all }i\right\} $. We call $V$ the
basis of the lattice $L(V)$. We say that a set of vectors $L
\subset {\cal R}^n$ is a lattice if there is a basis $V$ such that
$L = L(V)$.

Finding ``short vectors'' (i.e., vectors with small Euclidean
norm) in lattices is a hard computational problem. There are no
known efficient algorithms to find or even approximate - given an
arbitrary basis of a lattice - either the shortest non-zero vector
in the lattice, or another basis for the same lattice whose
longest vector is as short as possible. Given an arbitrary basis
$B$ of a lattice $L$ in $\R^n$, the best algorithm to approximate
(up to a polynomial factor in $n$) the length of the shortest
vector in $L$ is the $L^3$ algorithm \cite{LLL} which approximates
these problems to within a ratio of $2^{n/2}$ in the worst case,
and its improvement \cite{Sch} to ratio $(1+\e)^n$ for any fixed
$\e > 0$.

Ajtai reduced the worst-case complexity of problem (W) which is
closely related the length of the shortest vector and basis in a
lattice, to the average-case complexity of problem (A) (version
presented here is due to Goldreich, Goldwasser, and Halevi
\cite{GGH}).

\begin{itemize}
\item[W]: Given an arbitrary basis $B$ of a lattice $L$,
find a set of $n$ linearly independent lattice vectors, whose
length is at most polynomially (in $n$) larger than the length of
the smallest set of $n$ linearly independent lattice vectors. (The
length of a set of vectors is the length of its longest vector.)

\item[A]: Let parameters $n, m, q\in {\cal N}$ be such that
$n\log q < m \leq \frac{q}{2n^4}$ and $q = O(n^c)$ for some
constant $c > 0$. Given a matrix $M \in \Z_q^{n \times m}$, find a
vector $x \in \{-1, 0, 1\}^m, x \neq 0$ so that $M x \equiv 0
\pmod{q}$.
\end{itemize}

\begin{thm}{\rm \cite{Aj,GGH}}
Suppose that it is possible to solve a uniformly selected instance
of Problem~(A) in expected $T(n,m,q)$-time, where the expectation
is taken over the choice of the instance as well as the
coin-tosses of the solving algorithm. Then it is possible to solve
Problem~(W) in expected $\poly(|I|)\cdot T(n,\poly(n),\poly(n))$
time {\em on every} $n$-dimensional instance $I$, where the
expectation is taken over the coin-tosses of the solving
algorithm.
\end{thm}

The construction of a candidate one-way function follows in a
straight forward fashion. Let $M$ be a random $k \times m$ matrix
$M$ with entries from $\Z_q$, where $m$ and $q$ are chosen so that
$k\log q < m < \frac{q}{2k^4}$ and $q = O(k^c)$ for some constant
$c > 0$ ($k$ here is the security parameter).

The one-way function candidate is then $f(M,s)= (M,  M s \bmod{q}
= \sum_i s_i M_i \bmod{q} )$ where $s = s_1 s_2 \cdots s_m \in
\{0,1\}^m$ and $M_i$ is the $i$'th column of $M$. We note that
this function is regular.

\subsubsection{Factoring integers proposal}

\vskip-5mm \hspace{5mm}

Consider the function $Squaring(n,x)=(n,x^2\bmod n)$ where $n=pq$
for $p,q\in Z$ prime numbers and $x\in Z_n^*$, and the
corresponding collection of functions $Squaring=\{
Squaring_n(x)=x^2\bmod n : Z_n^*\rightarrow Z_n^*, n=pq, p,q\hbox{
primes}, |p|=|q|=k\}_k$. This function is easy to compute without
knowing the factorization of $n$, and is easy to invert given the
factorization of $n$ (the trapdoor) using fast square root
extraction algorithms modulo prime moduli \cite{Angluin82} and the
Chinese remainder theorem. Moreover, as the primes are abundant by
the prime number theorem ($\approx {1\over k}$ for $k$-bit primes)
and there exist probabilistic expected polynomial time algorithms
for primality testing \cite{GK86,AH}, it is easy to uniformly
select $n,p,q$ of the right form.

In terms of hardness to invert, Rabin \cite{Ra} has shown it as
hard to invert as it is to factor $n$ as follows. Suppose there
exists a factoring algorithm $A$. Choose $r\in Z_n^*$ at random.
Let $y= A(r^2\bmod n)$. If $y \neq r$ or $n-r$, then let
$p=gcd(r-y, n)$, else choose another $r$ and repeat. Within
expected 2 trials you should obtain $p$. The asymptotically proven
fastest integer factorization algorithm to date is the number
field sieve which runs in expected time $e^{((c + o(1))(\log
n)^{\frac{1}{3}}(\log \log n)^{\frac{2}{3}})}$ \cite{Pollard}. The
hardest input to any factoring algorithms are integers $n=pq$
which are product of two primes of similar length. Finally, for a
fixed $n$, $Squaring(n,\cdot )$ can be shown as hard to invert on
the average over $x\in Z_n^*$ as it is for any $x$. We remark,
that integer factorization has been first proposed as a basis for
a trapdoor function in the celebrated work of Rivest, Shamir and
Adelman \cite{RSA}.

By choosing $p$ and $q$ to be both congruent to $3\bmod 4$ and
restricting the domain of $Squaring_n$ to the quadratic residues
mod $n$, this collection of functions becomes a collection of
permutations proposed by Williams \cite{Wi}, which are especially
easy to work with in many cryptographic applications.

An open problem is to prove that the difficulty of factoring
integers is as hard on the average as in the worst case. In our
terminology an affirmative answer would mean that $x^2\bmod n$ is
as hard to invert on the {\it average} over $n$ and $x$, as it is
for {\it any}  $n$ and $x$.

\subsubsection{Quadratic residues vs. quadratic non residues proposal}

\vskip-5mm \hspace{5mm}

Let $n\in Z$. Then we call $y\in Z_n^*$ is a {\it quadratic
residue mod} $n$ iff $\exists x\in Z_n^*$ such that $y\equiv x^2
\bmod n$. Let us restrict our attention to $n=pq$ where
$p=q=3\bmod 4$.

Selecting a random quadratic residue mod $n$ is easy by choosing
$r \in Z_n^*$ and computing $r^2\bmod n$. Similarily, for such
$n$, selecting a random quadratic non-residue is easy by choosing
$r \in Z_n^*$ and computing $n-r^2\bmod n$ (this is a quadratic
non-residue by the property of the $n$'s chosen).

On the other hand, deciding whether $x$ is a quadratic residue
modulo $n$ for $n$ composite (which is the case if and only if it
is a quadratic residue modulo each of its prime factors), seems a
hard computational problem. No algorithm is known other than first
factoring $n$ and then deciding whether $x$ is a quadratic residue
modulo all its prime factors. This is easy for a prime modulos by
computing the Legendre symbol ${({{x}\over {p}})} =
{x^{{p-1}\over2}}\bmod p$ ($=1$ iff $x$ is a quadratic residue mod
$p$). The Legendre symbol is generalizable to the Jacobi symbol
for composite moduli ${({x\over n })} = {\Pi_{{p^{\alpha}}|n}{
{({ x\over{p}} )}^{\alpha}}}$ where $n={\Pi{p^{\alpha}}}$. The
Jacobi symbol only provides partial answer to whether $x\bmod n$
is a quadratic residue or not. For $x\in {J_n^{+1}} = \{ x\in
Z_n^*, {({x\over n} )} =1 \}$, it gives no information.

A proposal by Goldwasser and Micali \cite{GM82} for a collection
of trapdoor predicates follows.

$QR= \{ QR_{n}: J_n^{+1}\rightarrow\{ 0,1 \}\}_{n\in I}$ where
$I=\{n=pq \| p,q, \hbox{ primes}, |p|=|q|\}$,
\[ QR_{n}(x) = \left\{ \begin{array}{l}
\mbox{0 if $x$ is a quadratic residue mod $n$} \\
\mbox{1 if $x$ is a quadratic non-residue mod $n$}
\end{array} \right\}. \]

It can be proved that for every $n$ distinguishing between random
quadratic residues and random quadratic non residues with Jacobi
symbol +1, is as hard as solving the problem entirely in the worst
case.

\begin{thm}{\rm \cite{GM82}}  Let $S\subset I$.
If there exists a PPT algorithm which for every $n\in S$, can
distinguish between quadratic residues and quadratic non-residues
with non-negligible probability over $1\over 2$ (probability taken
over the $x\in Z_n^*$ and the coin tosses of the distinguishing
algorithm), then there exist a PPT algorithm which for every $n\in
S$ and every $x\in Z_n^*$ decides whether $x$ is a quadratic
residue mod $n$ with probability close to 1.
\end{thm}

\section{Encryption case study}

\vskip-5mm \hspace{5mm}

As discussed in the introduction we would like to propose
cryptographic schemes for which we can prove theorems guaranteeing
the security of our proposals. This task includes a definition
phase, construction phase and a reduction proof which is best
illustrated with an example. We choose the example of encryption.

We will address here the simplest setting of a passive adversary
who can tap the public communication channels between
communicating parties. We will measure the running time of the
encryption, decryption, and adversary algorithms as a function of
a  {\it security parameter} $k$ which is a parameter fixed at the
time the cryptosystem is setup.  We model the adversary as any
probabilistic algorithm which runs in time bounded by some
polynomial in $k$. Similarily, the encryption and decryption
algorithms designed are probabilistic and run in polynomial time
in $k$.

\subsection{Encryption: definition phase}

\vskip-5mm \hspace{5mm}

\begin{defn}
\label{encrypt-public.def} A {\em public-key encryption scheme} is
a triple, $(G,E,D)$, of probabilistic polynomial-time algorithms
satisfying the following conditions
\begin{enumerate}
\item key generation algorithm :
On  input $1^k$ (the security parameter) algorithm $G$, produces a
pair $(e,d)$ where $e$ is called the public key, and $d$ the
corresponding private key. (Notation: $(e,d) \in {G}(1^k)$.) We
will also refer to the pair $(e,d)$ a pair of {\em
encryption}/{\em decryption} keys.
\item An encryption algorithm:
Algorithm $E$ takes as inputs encryption key $e$ from the range of
$ G(1^k)$  and string $m\in \{0,1\}^k$ called the {\it message},
and produces as output  string $c\in \{0,1\}^*$ called the {\it
ciphertext}. (We use the notation $c\in E(e,m)$ or the shorthand
$c\in E_e(m)$.) Note that as $E$ is probabilistic, it may produce
many ciphertexts per message.
\item A decryption algorithm:
Algorithm $D$ takes as input decryption key $d$ from the range of
$G(1^k)$, and a ciphertext  $c$ from the range of $E(e,m)$, and
produces as output a string $m'\in\{0,1\}^*$, such that for every
pair $(e,d)$ in the range of $G(1^k)$, for every $m$, for every
$c\in E(e,m)$, the $prob(D(d,c) \neq m')$ is negligible.
\item Furthermore, this system is ``secure''
(see discussion below ).
\end{enumerate}
\end{defn}

A {\em private-key encryption scheme} is identically defined
except that $e=d$. The security definition for private-key
encryption  and public-key encryption are different in one aspect
only, in the latter $e$ is a public input available to the whereas
in the former $e$ is a secret not available to the adversary.

\subsubsection{Defining security} \label{def:secure}

\vskip-5mm \hspace{5mm}

Brain storming about what it means to be secure brings immediately
to mind  several desirable properties. Let us start with the the
minimal requirement and build up.

First and foremost the private key should not be recoverable from
seeing the public key. Secondly, with high probability for any
message space, messages should not be entirely recovered from
seeing their encrypted form and the public file. Thirdly, we may
want that in fact no useful information can be computed about
messages from their encrypted form. Fourthly, we do not want the
adversary to be able to compute any useful facts about traffic of
messages, such as recognize that two messages of identical content
were sent, nor would we want her probability of successfully
deciphering a message to increase if the time of delivery or
relationship to previous encrypted messages were made known to
her.

In short, it would be desirable for the encryption scheme to be
the mathematical analogy of opaque envelopes containing a piece of
paper on which the message is written.  The envelopes should be
such that all legal senders can fill it, but only the legal
recipient can open it.

Two definitions of security attempting to capture the ``opaque
envelope'' analogy have been proposed in the work of \cite{GM82}
and are in use today: computational indistinguishability  and
semantic security. The first definition is easy to work with
whereas the second seems to be the natural extension of Shannon's
perfect secrecy definition to the computational world. They are
equivalent to each other as shown by \cite{GM82,Sloan}.

The first definition essentially requires that the the adversary
cannot find a pair of messages $m_0, m_1$ for which the
probability distributions over the corresponding ciphertexts is
computationally distinguishable.

\begin{defn}
We say that a {\it Public Key Cryptosystem $(G,E,D)$  is
computationally indistinguishable} if $\forall$ PPT algorithms
$F,A$, and for $\forall$ constant $c>0$, $\exists k_0$, $\forall$
$k>k_0$, $\forall  m_0,m_1\in F(1^k) $, $|m_0|=|m_1|$,
\begin{eqnarray*}
  & & |\Pr[A(e,c) = 1 \mbox{ where } (e,d)\in G(1^k); \, c\in E(e,m_0)] \\
  & & - \Pr[A(e,c) = 1  (e,d)\in G(1^k);\, c\in E(e,m_1)]| <
  \frac{1}{k^c}.
\end{eqnarray*}
\end{defn}

\noindent{\bf Remarks about the definition}
\begin{enumerate}
\item
In the case of private-key cryptosystem, the definition changes
slightly. The encryption key $e$ is not given to algorithm $A$.
\item
Note that even if the adversary know that the messages being
encrypted is one of two, he still cannot tell the distributions of
ciphertext of one message apart from the other.
\item
Any cryptosystem in which the encryption algorithm $E$ is
deterministic immediately fails to pass this security requirement.
(e.g given $e, m_0,m_1$ and $c$ it would be trivial to decide
whether $c=E(e,m_0)$ or $c=E(e,m_1 )$ as for each message the
ciphertext is unique.)
\end{enumerate}

The next definition is called {\it Semantic Security}. It may be
viewed as a computational version of Shannon's perfect secrecy
definition. It requires that the adversary should not gain any
computational advantage or partial information from having seen
the ciphertext.

\begin{defn}
We say that an {\it public key cryptosystem $(G,E,D)$ is
semantically secure} if $\forall$ PPT algorithm $A$ $\exists$ PPT
algorithm $B$, s.t. $\forall$ PPT algorithm $M$, $\forall$
function $h: M(1^k)\rightarrow\{0,1\}^*$, $\forall c>0$, $\exists
k_0$, $\forall k>k_0$, $\Pr[A(e,|m|,c) = h(m) \mid (e,d)\in
G(1^k)\next m\in M(1^k)\next c\in E(e,m)] \leq \Pr[B(e,|m|)=h(m)
\mid m\in M(1^k) ] + {1\over{k^c}}.$
\end{defn}

The algorithm $M$ corresponds to the message space from which
messages are drawn, and the function $h(m)$ corresponds to
information about message $m$ ( for example, $h(m)=1$ if $m$ has
the letter `e' in it).

\begin{thm}{\rm \cite{GM82,Sloan}}
A Public Key Cryptosystem is computationally  indistinguishable if
and only if it is semantically secure.
\end{thm}

\subsection{Encryption: construction phase}

\vskip-5mm \hspace{5mm}

We turn now to showing how to actually build a public key
encryption scheme which is polynomial time indistinguishable.  The
construction shown here is by Goldwasser and Micali \cite{GM82}.
The key to the construction is to answer a simpler problem: how to
securely encrypt single bits. Encrypting general messages would
follow by viewing each message as a string of bits each encrypted
independently.

Given a collection of trapdoor predicates B, we define a public
key cryptosystem $(G,E,D)_B$ as follows:

\begin{defn}
\label{encrypt.def} \noindent A probabilistic encryption $PE_B
=(G,E,D)$ based on trapdoor predicates $B$ is defined as:

\begin{enumerate}
\item Key generation algorithm G:
On input $1^k$, G outputs $(i,t_i)$ where $B_i\in B$,
$i\in\{0,1\}^k$ and $t_i$ is the trapdoor information. The public
encryption key is $i$ and the private decryption key is $t_i$.
(This is achieved by running the sampling algorithm $S_1$ from the
def of B.)
\item  Let $m = m_1 \ldots m_n$ where $m_j \in \{0,1\}$ be the
 message.
\begin{tabbing}
$E(i,m)$ encrypts $m$ as follows: \\
\hspace*{5mm} \= Choose $x_j \in_R D_i$ such that $B_i(x_j) = m_j$
                                                for $j = 1, \ldots, n$. \\
                \> Output $c = f_i(x_1) \ldots f_i(x_n)$. \\
\end{tabbing}
\vskip-5mm \hspace{5mm }
\item Let $c = y_1 \dots y_k$ where $y_i \in D_i$ be the cyph
ertext.
\begin{tabbing}
$D(t_i,c)$ decrypts $c$ as follows: \\
\hspace*{5mm} \= Compute $m_j = B_i(y_j)$ for $j = 1, \ldots, n$.
\\
                \> Output $m = m_1 \ldots m_n$. \\
\end{tabbing}
\end{enumerate}

\end{defn}

It is clear that all of the above operations can be done in
expected polynomial time from the definition of trapdoor
predicates and that messages can indeed be sent this way.

Let us ignore for a minute the apparent inefficiency of this
proposal in bandwidth expansion and computation (which has been
addressed by Blum and Goldwasser in \cite{BG}) and talk about
security. It follows essentially verbatim from the definition of
trapdoor predicates that this system is polynomially time
indistinguishable in the case the message is a single bit (i.e.
$n=1$). Even though every bit individually is secure, it is
possible in principle that some predicate computed on all the bits
(e.g. their parity) is easily computable. Luckily, it is not the
case.

We prove polynomial time indistinguishability using the {\it
hybrid argument}. This method is a key proof technique in the
theory of pseudo randomness and secure protocol design, in
enabling to show how to convert a slight ``edge'' in solving a
problem into a complete surrender of the problem.

As this is one of the most straight forward simplest examples of
this technique we shall give it in full.

\begin{thm}{\rm \cite{GM82}}
Probabilistic encryption $PE_B= (G,E,D)$ is semantically secure if
and only if $B$ is a collection of trapdoor predicates.
\end{thm}

\begin{proof}
Suppose that $(G,E,D)$ is not indistinguishably secure (i.e. not
semantically secure).  Then there is a $c>0$, a PPT $A$ and $M$
such that for infinitely many $k$, $\exists m_0, m_1 \in M(1^k)$
with $|m_0 |=|m_1 |$,
\begin{eqnarray*}
  (*) & & \Pr[A(i,c) = 1 \mbox{ where } (i,t_i)\in G(1^k); \, c \in E(i,m_0)]\\
  & &-\Pr[A(i,c) = 1  (i,t_i)\in G(1^k); \, c\in E(i,m_1)] \geq
  \frac{1}{k^c},
\end{eqnarray*}
where the probability is taken the choice of $(i,t_i)$, the coin
tosses of $A$ and $E$.

Consider  $k$ where (*) holds. Wlog, assume that $|m_0 |=|m_1 |=k$
and that $A$ says $0$ more often when $c$ is an encryption of
$m_0$ and $1$ more often when $c$ is an encryption of $m_1$.

Define distributions $D_j = E(i,s_j)$ for $j = 0,1, \ldots, k$
where $s_0 = m_0, s_k = m_1$ and $s_j$ differs from $s_{j+1}$ in
precisely $1$ bit.

Let $P_j = \Pr[A(i,c) = 1 | c \in D_j]$.

Then $P_k - P_0 \geq  \frac{1}{k^c}$ and since
$\sum_{j=0}^{k-1}(P_{j+1} - P_j) = P_k - P_0$, $\exists j$ such
that $P_{j+1} - P_j \geq  \frac{1}{k^{c+1}}$.

Assume that $s_j$ and $s_{j+1}$ differ in the $l^{\mbox{\tiny
th}}$ bit; that is, $s_{j,l} \neq s_{j+1,l}$ or, equivalently,
$s_{{j+1},l} = \bar{s_{j,l}}$ where $s_{j,u}$ is the $u$-th bit of
$s_j$.

Now, consider the following algorithm $B$ which takes input $i,y$
and outputs $0$ or $1$ as its guess to the value of the hard core
predicate $B_i(y)$.

$B$ on input $i,y$:
\begin{enumerate}
\item  Choose $y_1, \ldots, y_k$ such that $B_i(y_r) = s_{j,r}$
for $r = 1, \dots, k$ using $S_1$ from the definition of $B$.
\item  Let $c = y_1, \ldots, y, \ldots, y_k$
where $y$ has replaced $y_l$ in the $l^{\mbox{\tiny th}}$ block.
\item  If $A(1^k,i,,m_0,m_1,c) = 0$ then output $s_{j,l}$. \\
If $A(1^k,i,,m_0,m_1,c) = 0$ then output $s_{j+1,l} =
\bar{s}_{j,l}$.
\end{enumerate}
Note that $c \in E(i,s_j)$ if $B_i(y) = s_{j,l}$ and $c \in
E(i,s_{j+1})$ if $B_i(y) = s_{j+1,l}$.

Thus, in step $3$ of algorithm $B$, outputting $s_{j,l}$
corresponds to $A$ predicting that $c$ is an encryption of $s_j$.
\medskip

\noindent {\bf Claim}\ \ {$\Pr[B(i,y) = B_i(y)] > \frac{1}{2} +
\frac{1}{k^{c+1}}$}.

\begin{proof}
\begin{eqnarray*}
  \Pr[B(i,f_i(y)) = B_i(y)] & = & \Pr[A(i,c) = 0 | c \in E(i,s_j)]\Pr[c \in E(i,s_j)] \\
  & & + \Pr[A(i,c) = 1 | c \in E(i,s_{j+1})] \Pr[c \in E(i,s_{j+1})] \\
  & \geq & (1-P_j)(\frac{1}{2}) + (P_{j+1})(\frac{1}{2}) \\
  & = & \frac{1}{2} + \frac{1}{2}(P_{j+1} - P_j) \\
  & > & \frac{1}{2} + \frac{1}{k^{c+1}}.
\end{eqnarray*}

Thus, $B$ will predict $B_i(y)$  given $i$, $y$ with probability
better than $\frac{1}{2} + \frac{1}{k^{c+1}}$.  This contradicts
the assumption that $B_i$ is a trapdoor predicate.
\end{proof}

Hence, the probabilistic encryption $PE =(G,E,D)$ is
indistinguishably secure.
\end{proof}

\subsection{Strengthening the adversary: non malleable security}

\vskip-5mm \hspace{5mm}

The entire discussion so far has assumed that the adversary can
listen to the cipher texts being exchanged over the insecure
channel, read the public-file (in the case of public-key
cryptography), generate encryptions of any message on his own (for
the case of public-key encryption), and perform probabilistic
polynomial time computation.

One may imagine a more powerful adversary who can intercept
messages being transmitted from sender to receiver and either stop
their delivery all together or alter them in some way. Even worse,
suppose the adversary can after seeing a ciphertext, request a
polynomial number of related ciphertexts to be decrypted for him.
For definitions and constructions of encryption schemes secure
against such adverdary see \cite{SR,DDN, BFM,CS}.

%

\section{A constructive theory of pseudo randomness }

\vskip-5mm \hspace{5mm}

A theory of randomness based on computability theory was developed
by Kolmogorov, Solomonov and Chaitin \cite{Sol, Kol, Cha}.  This
theory applies to individual strings and defines the complexity of
strings as the shortest program (running on a universal machine)
that generates that string. A perfectly random string is the
extreme case for which no shorter program than the length of the
string itself can generate it. Inherintly, it is impossible to
generate perfect random strings from shorter ones.

One of the surprising contributions of cryptographically motivated
research in the early eighties, has been a theory of randomness
computational complexity theory pioneered by Shamir
\cite{Shamir81} Blum and Micali \cite{BM82}, which makes it
possible in principle to deterministically generate random strings
from shorter ones. Not to mix notions, we will henceforth refer to
this latter development as a theory of pseudo randomness, and the
strings generated as pseudo random. In contrast, when we speak of
choosing a truly random string of a fixed length over some
alphabet, we refer to selecting it with uniform probability over
all strings of the same length. In this section we shall only
speak of binary alphabet. The notation $x\in_R \{0,1\}^k$  will
thus be taken to mean that for every $s\in \{0,1\}^k$, the
probability of $x=s$ is $1/2^k$.

Defining pseudo-random distributions is a special case of the
definition of computational indistinguishability, which we
encountered earlier in the context of secure encryption. A
distribution over binary strings is called {\it pseudo-random} if
it is computationally indistinguishable from the uniform
distribution over all binary strings of the same length. The idea
is that as long as we cannot tell apart samples from the uniform
distribution from samples of a distribution $X$ in polynomial time
, there is no difference between using either distributions that
can be observed in polynomial time. In particular, any
probabilistic algorithm, in which the internal coin flips of the
algorithm are replaced by strings sampled from $X$, must not
behave any different than it would using truly random coin flips.
A counter example will yield a statistical test to distinguish
between $X$ and the uniform distribution.

A deterministic polynomial time program which 'stretchs' a short
input string selected with uniform distribution (henceforth called
the `seed'), to a polynomial long output string is called a pseudo
random sequence generator. When such a construction is accompanied
with a proof that the output string distribution is pseudo random
we call the generator a strong pseudo random sequence generator
(SPRSG).\footnote{ Again the choice of polynomial-time is
arbitrary here, a strong pseudo random sequence generator can be
defined to be a deterministic program which works in time $T(n)$
where $n$ is the seed length and is computationally
indistinguishable with respect to algorithms which run in time
$T'(n)$ for time functions $T,T'$.}

In a culmination of a sequence of results  by
\cite{Shamir81,BM82,Y82,GGM,hill}, Hastand, Impagliazzo, Levin and
Luby showed that a necessary and sufficient condition for the
existence of strong pseudo random sequence generators is the
existence of one-way functions.

The link between one-way functions and pseudo randomness starts
from the following observation. First, rephrase the fact that
inverting one-way functions is difficult, by saying that the
inverse of a one-way function is unpredictable. In particular, the
hard-core of a one-way function is impossible to predict with any
non-negligible probability greater than $1\over 2$. Second, show
that impossibility to predict is the ultimate test for pseudo
randomness. Namely, if a pseudo-random sequence generator has the
property that it is difficult to predict the next bit from
previous ones with probability significantly better than $1\over
2$ in time polynomial in the size of the seed, then it is
impossible to distinguish in polynomial time between strings
produced by the pseudo random sequence generators and truly random
strings. This is proved by turning any statistical test that
distinguishes in polynomial time pseudo random strings from random
strings into polynomial time next bit predictor. This link is not
conditional on the existence of one-way functions. In fact, in
work by Nisan and Wigderson \cite{NW} they removed the requirement
that the pseudo random sequence generator has to work in time
which is as fast as the algorithm trying to distinguish the output
sequences from truly random. Generators of this type are generally
useless for cryptographic applications (as they can not be
generated in feasible time) but are very useful for proving
complexity theoretic results.

Strong pseudo random generators are useful for understanding the
relation between deterministic algorithms and probablistic
algorithms. The idea which was put forth by Yao \cite{Y82} was to
replace a single execution of a probablistic polynomial time
algorithm $A$ with the majority output of all the executions of
the same algorithm, where each execution uses instead of random
coins the output of a strong pseudo random number generator on a
different input seed. The cost of the latter deterministic
procedure will be a factor of $2^{k'}$ longer where $k'$ is the
seed length used to generate the pseudo random sequences
necessary. The algorithm $A$ must behave ``the same'' when it uses
truly random coins as when it uses coins which are pseudo-random,
as otherwise it becomes a distinguisher between the uniform and
pseudo-random distributions, an impossible task for a
probabilistic polynomial time algorithm. Putting this together, we
get : if one-way functions exist, then $BPP \subseteq
\cap_{\epsilon} DTIME(2^{k^{\epsilon}}) $. This tradeoff between
the {\it hardness} of inverting the one-way function, and {\it
randomness} replacement, has been followed up with many papers in
complexity theory each either relaxing the hardness assumption or
tightening the relation between deterministic and probabilistic
complexity classes.

Strong pseudo random generators are particularly useful for
cryptography. Suppose you need a large supply of random strings
for your cryptographic applications (e.g. the choice of secret
keys, internal coin tosses of an encryption algorithm, etc.). If
you use instead of truly random bits, pseudo random sequence
generators which are weak (e.g. predictable), it may completely
destroy the underlying cryptographic applications \cite{BGM}. In
contrast, we  can replace any use of truly random coins with
strong pseudo random ones (assuming we have access to truly random
coins for the seeds --- which is an interesting discussion all by
itself), without fear of compromising the security of the
underlying application. Indeed, if as a result of such replacement
the cryptographic application becomes insecure, then a way is
found to distinguish outputs of SPRG from the uniform
distribution. Many classical pseudo random number generators which
are quite useful and effective for Monte Carlo simulations, have
been shown not only weak but predictable in a strong sense which
makes them typically unsuitable for cryptographic applications.
For example, {\em linear} feedback shift registers \cite{Golomb82}
are well-known to be cryptographically insecure; one can solve for
the feedback pattern given a small number of output bits, and
similarily outputs of linear congruential generators
\cite{FriezeHaKaLaSh88}. In \cite{KannanLeLo84} Kannan, Lenstra,
and  Lovasz use the $L^3$ algorithm to show that the binary
expansion of any algebraic number $y$ (such as
$\sqrt{5}=10.001111000110111\ldots$) is insecure, since an
adversary can identify $y$ exactly from a sufficient number of
bits, and then extrapolate $y$'s expansion.

\subsection{Pseudo random functions, permutations, and what else?}

\vskip-5mm \hspace{5mm}

Similarily to defining pseudo random sequences one may ask what
other random objects can be replaced with pseudo-random counter
parts. Goldreich, Goldwasser and Micali \cite{GGM} considered in
this light random functions, which from a gold mind for
applications. Pseudo random functions are defined to be for every
size $k$ a subset of all functions from (and to) the binary
strings of length $k$, which are polynomial time indistinguishable
from truly random functions by any algorithm whose only access to
the function is to query it on inputs of its choice. However, in
contrast with a truly random function, a pseudo random function
has a short description which if known enables efficient
evaluation.

Let $H_k = \{f:\{0,1\}^{k} \rightarrow \{0,1\}^{k}\}$ then $|H_k|
= (2^{k})^{2^{k}}$.  Let ${\cal H} = \bigcup_{k} H_k$.

\begin{definition}
A {\em polynomial time statistical test for functions} is a
polynomial time algorithm $T^{f}$ with access to a black box $f$
from which $T$ can request values of $f(x)$ for x of $T$'s choice.
A collection of functions ${\cal F} = \bigcup_k F_k$ where $F_k
\subset H_k$ {\em passes the statistical test $T$} if $\forall Q
\in {\bf Q}[x], \exists k_0, \forall k>k_{0} \: |T(F_k) - T(H_k)|
< \frac{1}{Q(k)}$ where $T(F_k) = \Pr_{f \in F_k,
\mbox{\scriptsize coins of T}}[T^{f}(1^k) = 1]$ and $T(H_k) =
\Pr_{f \in H_k, \mbox{\scriptsize coins of} \: T}[T^{f}(1^k) =
1]$.
\end{definition}

\begin{definition}
A collection of functions ${\cal F} = \bigcup_k F_k$ is a {\em
pseudo-random collection of functions} if
\begin{enumerate}
\item (Indexing) For each k, there is a unique index $i \in \{0,1\}^k$
associated with each $f \in F_k$. The function $f \in F_k$
associated with index $i$ will be written $f_{i}$.

\item (Efficiency) There is a polynomial time function $A$ so that
$A(i,x) = f_i(x)$.

\item (Pseudo-randomness) ${\cal F}$ passes all polynomial time
statistical tests for functions.
\end{enumerate}
\end{definition}

\begin{theorem}{\rm \cite{GGM}}
If there exist one-way functions, then there exist pseudo-random
collections of functions.
\end{theorem}

An immediate application of pseudo random functions is the
construction of semantically secure private key cryptosystem  as
follows. Let $s$ an index of a pseudo random function $f_s$ be the
joint secret key of the sender Alice and the receiver Bob. Then to
encrypt message $m$, Alice selects at random $r\in\{0,1\}^k$, and
sets the cipher text $c=(r,f_s(r)\xor m)$ where $\xor$ is the
bit-wise exclusive-or of two strings. To decrypt $c=(a,b)$, Bob
computes $f_s(a)\xor b$.

Pseudo random functions have been used to derive negative results
in computational learning theory by Valiant and Kearns \cite{VK}.
They show that any concept class (i.e. a set of Boolean functions)
which contains a family of pseudo random functions cannot be
efficiently learnable under the uniform distribution and with the
help of membership queries. A learning algorithm is given oracle
access to any function in the class and is required to output a
description of a function which is close to the target function
(being queried).

The work on {\it natural proofs} originated by Rudich and Razborov
\cite{RR} use pseudo random functions to derive negative results
on the possibility of proving good complexity lower bounds using a
restricted class of circuit lower bound proofs referred to as {\it
natural}. It is proved that natural (lower bound) proofs cannot be
established for complexity classes containing a family of pseudo
random functions.

An interesting question is to characterize which classes of random
objects can be replaced by pseudo random objects. Luby and Rackoff
\cite{LR} treated the case of pseudo random permutations and Naor
and Reingold the case of permutations with cyclic structure
\cite{NR}. As any object can be abstracted as a restricted class
of functions, the real question is what form of access to the
function does the statistical test have. In the standard
definition, the statistical test for functions can query the
functions at values of its choice. This may not be necessarily the
natural choice in every case. For example, if the function
corresponds to the description of a random graph (e.g. $f(u,v)=1$
if and only if an edge is present between vertices $u$ and $v$).

Define the ``ultimate'' extension of a statistical test for
functions on $k$ bit strings, to be given access to the {\it
entire truth table} of the function (i.e. an exponential size
input). The following observation is then straightforward.

\begin{thm}
Let $f:\{0,1\}^*\rightarrow \{0,1\}^*$ be polynomial time
computable function, for which the fastest inverting algorithm
runs in time $2^{n^\epsilon}$ for some $\epsilon > 0$. Then, there
exist collections of pseudo random functions which pass all
ultimate statistical tests for functions.
\end{thm}

\section{Interactive protocols, interactive proofs, and  zero
knowledge interactive proofs}

\vskip-5mm \hspace{5mm}

Secure one-way communication is a special case of general
interactive protocols. The most exciting developments in
cryptography beyond public-key cryptography has been the
development of interactive protocols, interactive proofs, and zero
knowledge interactive proofs \cite{GMR86,GMW1, Y82, GMW2, BGW88,
CCD88, Babai}. \footnote{In particular, the idea of multi-prover
interactive proofs of Benor, Goldwasser, Kilian, and Wigderson
\cite{BGKW} (which has become better known as {\it probabilistic
checkable proofs}) has led to a rice body of NP-hardness results
for approximation versions of optimization problems
\cite{Johnson}.} Unfortunately, we have no space to cover these
developments in this article.  These topics have been surveyed
extensively, and the interested reader may turn to
\cite{GO1,book}.

A few final words. Generally speaking, an {\it interactive
protocol} consists of  two or more parties who cooperate and
coordinate without a trusted ``third'' party to accomplish a
common goal,  referred to as the {\it functionality} of the
protocol, while maintaining the {\it secrecy} of their private
data. A functionality may be computing a simple deterministic
function such as majority of the inputs of the communicating
parties, or a more complicated probabilistic computation such as
playing a non-cooperative game without a trusted referee.

In the case of more than two parties, the case of adversarial
coalitions of participants  who attempt to damage the
functionality and break secrecy has been considered. Very powerful
and surprising theorems about the ability of playing
non-cooperative games without a trusted ``third party'' have been
shown. A sample theorem of Benor, Goldwasser, and Wigderson shows
that in the presence of an adversarial coalition containing less
than a third of the parties, any probabilistic computation can be
performed maintaining functionality and perfect information
theoretic secrecy of the inputs, as long as each pair of parties
can communicate in perfect secrecy \cite{BGW88,CCD88}. These
results make extensive use of error correcting codes based on
polynomials. The connection between these theorems and research in
game theory and threory of auctions is well worth examining.


\label{lastpage}


\begin{thebibliography}{aa}

\bibitem{ACGS88}
W.~B. Alexi, B.~Chor, O.~Goldreich, and C.~P. Schnorr.
\newblock {RSA}/{Rabin} functions: certain parts are as hard as the whole.
\newblock {\em SIAM J. Computing}, 17(2):194--209, April 1988.

\bibitem{Adleman94}
L.~M. Adleman.
\newblock Algorithmic Number Theory --- The Complexity Contribution
\newblock Proceedings of the Foundations of Computer Science,
88--99, October 1994.

\bibitem{AH}
L.~M. Adleman and M.~A. Huang.
\newblock Recognizing primes in random polynomial time.
\newblock In {\em Proc.\ $19$th ACM Symp.\ on Theory of Computing},
462--469, New York City, 1987. ACM.

\bibitem{Aj} M.~Ajtai.
\newblock Generating Hard Instances of Lattice Problems.
\newblock In {\em 28th STOC}, 99--108, 1996.

\bibitem{Angluin82}
D.~Angluin.
\newblock Lecture notes on the complexity of some problems in number theory.
\newblock Technical Report TR-243, Yale University Computer Science Department, August 1982.

\bibitem{AL}
D. Angluin, D. Lichtenstein,
\newblock Provable security of cryptosystems:A survey.
\newblock Tech. Rep. 288, Dept. of Computer Science, Yale Univ. New Haven, Conn., 1983.

\bibitem{Bach88}
Eric Bach.
\newblock How to generate factored random numbers.
\newblock {\em SIAM J. Computing}, 17(2):179--193, April 1988.

\bibitem{BGW88}
M.~Ben-Or, S.~Goldwasser, and A.~Wigderson.
\newblock Completeness theorems for fault-tolerant distributed computing.
\newblock In {\em Proc.\ $20$th ACM Symp.\ on Theory of Computing},
1--10, Chicago, 1988. ACM.

\bibitem{Bl}
M.~Blum.
\newblock Coin flipping by telephone.
\newblock In {\em Proc. IEEE Spring COMPCOM}, 133--137. IEEE, 1982.

\bibitem{BM82}
M.~Blum and S.~Micali.
\newblock
How to generate cryptographically strong sequences of
pseudo-random
  bits.
\newblock {\em SIAM J.\ Computing}, 13(4):850--863, November 1984.

\bibitem{BG} M.~Blum and S.~Goldwasser.
\newblock An Efficient Probabilistic Public-Key Encryption Scheme
which hides all partial information.
\newblock In {\em Crypto{84}}, LNCS (196) Springer-Verlag, 289--302.

\bibitem{BFM}
M. Blum and P. Feldman and S. Micali,
\newblock Proving Security Against Chosen Cyphertext Attacks,
\newblock In proceedings of  CRYPTO88, 256--268, 1988.

\bibitem{Babai} L.~Babai.
\newblock Trading Group Theory for Randomness.
\newblock In {\em 17th STOC}, 421--420, 1985.

\bibitem{BGM}
Bellare, M. and Goldwasser, S. and Micciancio, D.,
\newblock ``Pseudo-Random'' Number Generation within Cryptographic A
lgorithms: The {DSS} Case,
\newblock In proceedings of Crypto '97, 277--291, 1977.

\bibitem{BGKW}
M. Ben-Or,  S. Goldwasser,J. Kilian, and A. Wigderson,
\newblock  Multi-Prover Interactive Proof-Systems,
\newblock Proceedings of the Twentieth Annual ACM Symposium on Theory of Computing (1988), 113--131.

\bibitem{Cha}
C.J. Chaitin
\newblock On the Length of Programs for Computing Finite Binary Sequences.
\newblock Journal of the ACM, vol 13, 547--570, 1966.

\bibitem{CS}
R. Cramer, V. Shoup,
\newblock ``A Practical Public Key Cryptosystem Provably Secure against Adaptive Chosen Ciphertext Attack'',
\newblock Advances in Cryptology --- CRYPTO '98 Proceedings, 13--25, Springer-Verlag, 1998

\bibitem{MC97}
Christian Cachin and Ueli Maurer
\newblock Unconditional Security Against Memory-Bounded Adversaries
\newblock Advances in Cryptology --- CRYPTO '97, Lecture Notes in Computer Science, Springer-Verlag, vol. 1294, 292--306, 1997.

\bibitem{CCD88}
Crepeau D. Chaum and I. Damgard.
\newblock Multiparty unconditionally secure protocols.
\newblock In Proc. of 20th ACM Symp. on Theory of Computing, Chicago, 1988.

\bibitem{DH76}
W.~Diffie and M.~E. Hellman.
\newblock New directions in cryptography.
\newblock {\em IEEE Trans.\ Inform.\ Theory}, IT-22:644--654, November 1976.


\bibitem{DDN}
D.~Dolev, C.~Dwork, and M.~Naor.
\newblock Non-malleable cryptography.
\newblock In {\em Proc.\ $23$rd ACM Symp.\ on Theory of Computing},
542--552. ACM, 1991.


\bibitem{FriezeHaKaLaSh88}
A.~M. Frieze, J.~Hastad, R.~Kannan, J.~C. Lagarias, and A.~Shamir.
\newblock Reconstructing truncated integer variables satisfying linear congruences.
\newblock {\em SIAM J. Computing}, 17(2):262--280, April 1988.


\bibitem{GGM} O.~Goldreich, S.~Goldwasser, and S.~Micali.
\newblock How to Construct Random Functions.
\newblock {\em JACM}, Vol.~33, No.~4, 792--807, 1986.



\bibitem{G01}
O.~Goldreich.
\newblock {\em Foundations of Cryptography: Volume 1 -- Basic Tools}.
\newblock Cambridge University Press, 2001.

\bibitem{Go}
Oded Goldreich. newblock A Note on Computational
Indistinguishability.
\newblock Information Processing Letters 34(6): 277--281 (1990)

\bibitem{GILVZ}
Oded Goldreich, Russell Impagliazzo, Leonid A. Levin, Ramarathnam
Venkatesan, David Zuckerman.
\newblock Security Preserving Amplification of Hardness.
\newblock FOCS 1990: 318--326

\bibitem{GK89}
Oded Goldreich, Hugo Krawczyk.
\newblock Sparse Pseudorandom Distributions.
\newblock CRYPTO 1989: 113--127

\bibitem{GL}
O.~Goldreich and L.~Levin.
\newblock A hard-core predicate for all one-way functions.
\newblock Proc., ACM Symp. on Theory of Computing, 25--32, 1989.

\bibitem{GMRi88}
S.~Goldwasser, S.~Micali, and Ronald~L. Rivest.
\newblock A digital signature scheme secure against adaptive chosen-message
  attacks.
\newblock {\em SIAM J.\ Computing}, 17(2):281--308, April 1988.

\bibitem{GK86}
S.~Goldwasser and J.~Kilian.
\newblock Almost all primes can be quickly certified.
\newblock In {\em Proc.\ $18$th ACM Symp.\ on Theory of Computing},
  316--329, Berkeley, 1986. ACM.

\bibitem{GM82}
S.~Goldwasser and S.~Micali.
\newblock Probabilistic encryption.
\newblock {\em JCSS}, 28(2):270--299, April 1984.

\bibitem{GMR86}
S.~Goldwasser, S.~Micali, and C.~Rackoff.
\newblock The knowledge complexity of interactive proof-systems.
\newblock {\em SIAM. J. Computing}, 18(1):186--208, February 1989.

\bibitem{GL02}
S. Goldwasser, and Y. Lindell.
\newblock
\newblock Proceedings of 16th International Symposium on Distributed
Computing
\newblock to appear October 2002.

\bibitem{GGH}
Goldwasser, Goldreich, Halevi.
\newblock Collision-Free Hashing from Lattice Problems.
\newblock Electronic Colloquium on Computational Complexity (ECCC) 3(42): (1996)

\bibitem{GMW2} O.~Goldreich, S.~Micali and A.~Wigderson.
\newblock How to Play any Mental Game ---
A Completeness Theorem for Protocols with Honest Majority.
\newblock In {\em 19th STOC,} 218--229, 1987.

\bibitem{GMT}
S.~Goldwasser, S.~Micali, and P.~Tong.
\newblock Why and how to establish a private code on a public network.
\newblock In {\em Proc.\ $23$rd IEEE Symp.\ on Foundations of Comp.\ Science},
134--144, Chicago, 1982. IEEE.

\bibitem{Golomb82} S.~W. Golomb.
\newblock {\em Shift Register Sequences}.
\newblock Aegean Park Press, Laguna Hills, 1982.
\newblock Revised edition.


\bibitem{GMW1}
O.~Goldreich, S.~Micali, and A.~Wigderson.
\newblock Proofs that yield nothing but their validity and a methodology of
  cryptographic protocol design.
\newblock In {\em Proc.\ $27$th IEEE Symp.\ on Foundations of Comp.\ Science},
174--187, Toronto, 1986. IEEE.

\bibitem{GO1}
O. Goldreich,
\newblock The Foundations of Cryptography - Volume 1
\newblock ISBN 0-521-79172-3 Cambridge University Press.

\bibitem{book}
O. Goldreich
\newblock Modern Cryptography, Probabilistic Proofs and Pseudorandomness
\newblock ISBN 3-540-64766-x
Springer-Verlag, Algorithms and Combinatorics, Vol 17, 1998.

\bibitem{RMG} Y. Gertner, T. Malkin and O. Reingold.
\newblock On the impossibility of basing trapdoor functions on trapdoor predicates,
\newblock FOCS 2001.


\bibitem{hill}
J. Hastad, R. Impagliazzo, L. Levin, Michael Luby.
\newblock Construction of a pseudo-random generator from any one-way function.
\newblock SIAM J. Comput. 28(4):1364--1396, 1999.

\bibitem{Johnson}
D. Johnson
\newblock Tale of the Second Prover.
\newblock Journal of Algorithms. Vol 13.

\bibitem{KannanLeLo84}
R.~Kannan, A.~Lenstra, and L.~{Lov\'asz}.
\newblock Polynomial factorization and non-randomness of bits of algebraic and some transcendental numbers.
\newblock In {\em Proc.\ $16$th ACM Symp.\ on Theory of Computing}, 191--200, Washington, D.C., 1984. ACM.

\bibitem{Kilian88}
J.~Kilian.
\newblock Founding cryptography on oblivious transfer.
\newblock In {\em Proc.\ $20$th ACM Symp.\ on Theory of
Computing}.



\bibitem{koblitz}
N. Koblitz, A. Menezes, and S. Vanstone.
\newblock The state of elliptic curve cryptography.
\newblock Designs, Codes and Cryptography, 19 (2000), 173--193.

\bibitem{Kol}
A. Kolmogorov
\newblock Three approaches to the concept of the amount of information
\newblock Probl. of Inform. Trandm., Vol1/1/,1965.

\bibitem{LL90}
A.~K. Lenstra and H.~W. Lenstra, Jr.
\newblock Algorithms in number theory.
\newblock In Jan van Leeuwen, editor, {\em Handbook of Theoretical Computer
  Science (Volume A: Algorithms and Complexity)}, chapter~12, 673--715.
  Elsevier and MIT Press, 1990.


\bibitem{LLL} A.K.~Lenstra, H.W.~Lenstra, L.~Lov\'asz.
\newblock Factoring polynomials with rational coefficients.
\newblock {\em Mathematische Annalen} 261, 515--534 (1982).

\bibitem{LR} M. Luby and C. Rackoff,
\newblock Pseudo-Random Permutation Generators and Cryptographic Composition.
\newblock In proceedings of STOC86, 356--363, 1986.


\bibitem{Mau93}
Ueli Maurer,
\newblock Protocols for Secret Key Agreement by Public Discussion Based on Common Information
\newblock IEEE Trans. on Inform. Theory (1993).

\bibitem{NY89}
Naor and Yung,
\newblock Universal One-Way Hash Functions and
their Cryptographic Applications
\newblock Proceedings of the Twenty First Annual ACM Symposium on
Theory of Computing. (May 15--17 1989: Seattle, WA, USA)



\bibitem{Ni} Noam Nisan.
\newblock
Pseudorandom generators for space-bounded computation.
\newblock Proceedings of the Twenty Second Annual ACM Symposium on Theory of Computing, 204--212, Baltimore, Maryland, 14--16 May 1990.


\bibitem{odlyzko}
A. M. Odlyzko,
\newblock Discrete logarithms: The past and the future,
\newblock Designs, Codes, and Cryptography 19, 129--145, 2000

\bibitem{RR}
A.R. Razborov and S. Rudich.
\newblock Natural proofs.
\newblock Journal of Computer and System Science, Vol. 55 (1), 24--35, 1997.


\bibitem{RSA}
Ronald~L. Rivest, Adi Shamir, and Leonard~M. Adleman.
\newblock A method for obtaining digital signatures and public-key
  cryptosystems.
\newblock {\em Communications of the ACM}, 21(2):120--126, 1978.




\bibitem{NW} N. Nisan and A. Wigderson,
\newblock Hardness vs. Randomness
\newblock Journal of JCSS, Vol 49, No 2, 149--167, 1994.


\bibitem{NR}
M. Naor and O. Reingold,
\newblock Constructing pseudorandom permutations with a prescribed structure,
\newblock SODA 2001, 458--459, 2001.

\bibitem{Pollard}
A.~K. Lenstra, H.~W. Lenstra, Jr., M.~S. Manasse, and J.~M.
Pollard.
\newblock The number field sieve.
\newblock In {\em Proc.\ $22$nd ACM Symp.\ on Theory of Computing},
564--572, Baltimore, Maryland, 1990. ACM.

\bibitem{R90}
John Rompel.
\newblock One-way functions are necessary and sufficient for secure signatures.
\newblock Proceedings of the Twenty Second Annual ACM Symposium on Theory of Computing, 387--394, Baltimore, Maryland, 14--16 May 1990.

\bibitem{RB89}
T.~Rabin and M.~Ben-Or.
\newblock Verifiable secret sharing and multiparty protocols with honest
  majority.
\newblock In {\em 21st ACM Symposium on Theory of Computing}, 73--85,
  1989.

\bibitem{Ra}
M.~Rabin.
\newblock Digitalized signatures as intractable as factorization.
\newblock Technical Report MIT/LCS/TR-212, MIT Laboratory for Computer Science, January 1979.



\bibitem{Su}
Madhu Sudan.
\newblock Coding Theory: Tutorial and Survey.
\newblock Proceedings of the 42nd Annual Symposium on Foundations of Computer Science, 36--53, Las Vegas, Nevada, 14--17 October, 2001.

\bibitem{Sch} C.P.~Schnorr.
\newblock A hierarchy of polynomial time lattice basis reduction algorithms.
\newblock In {\em Theoretical Computer Science}, vol. 53, 1987,
201--224.

\bibitem{Shannon48}
C.~E. Shannon.
\newblock A mathematical theory of communication.
\newblock {\em Bell Sys.\ Tech.\ J.}, 27:623--656, 1948.

\bibitem{Shannon49}
C.~E. Shannon.
\newblock Communication theory of secrecy systems.
\newblock {\em Bell Sys.\ Tech.\ J.}, 28:657--715, 1949.



\bibitem{Sloan}
S.~Micali, C.~Rackoff, and R.~H. Sloan.
\newblock The notion of security for probabilistic cryptosystems.
\newblock {\em SIAM J.\ Computing}, 17(2):412--426, April 1988.

\bibitem{Sol}
R.J. Solomonoff
\newblock A formal theory of Inductive Inference.
\newblock Inform. and Control. Vol 7/1, 1--22, 1964.

\bibitem{SR}
Rackoff, C. and Simon, D. R.,
\newblock Non-interactive zero-knowledge proof of knowledge and chosen ciphertext attack,
YEAR   =        {1991},
\newblock Proceedings of Crypto '91, 433--444,




\bibitem{Shamir81} A.~Shamir.
\newblock On the generation of cryptographically strong pseudo-random sequences.
\newblock In {\em Proc.\ ICALP}, 544--550. Springer, 1981.



\bibitem{SA} A.~Sahai and S.~Vadhan. 
\newblock A Complete Promise Problem for Statistical Zero-Knowledge.
\newblock In {\em 38th FOCS}, 448--457, 1997.

\bibitem{Tr}
Luca Trevisan
\newblock Extractors and Pseudorandom Generators
\newblock J. of the ACM, 48(4):860--879, 2001.


\bibitem{VK}
M. Kearns, and L. Valiant.
\newblock Cryptographic limitations on learning Boolean formulae and finite automata,
\newblock J. Assoc. Comp. Mach., 41:1 (1994) 67--95.

\bibitem{Wi}
H. Williams,
\newblock ``A Modification of the RSA Public-Key Encryption Procedure'',
\newblock IEEE Trans. Information Theory, 26(6) (1980), 726--729.

\bibitem{W75}
A. D. Wyner,
\newblock The wire-tap channel.
\newblock Bell System Technical Journal, Vol. 54, no. 8, 1975, 1355--1387.


\bibitem{Y82}
A.~C. Yao.
\newblock Theory and application of trapdoor functions.
\newblock In {\em Proc.\ $23$rd IEEE Symp.\ on Foundations of Comp.\ Science},
80--91, Chicago, 1982. IEEE.
\end{thebibliography}
\end{document}